\documentclass[12pt]{aastex6}

\usepackage{graphicx,amssymb,amsmath,times}
\graphicspath{{../}{../PLW/}{../figures2/}{../temp/oldcatalogs/}{../temp/170708_catalogs/}}
\usepackage{longtable}
\usepackage{bm} 
\usepackage{url}
\usepackage[varg]{txfonts}
\usepackage[T1]{fontenc}
\usepackage[latin9]{inputenc}
\usepackage{natbib}
\usepackage{epstopdf}
\def\simlt{\lower.5ex\hbox{\ltsima}}
\def\simgt{\lower.5ex\hbox{\gtsima}}

\def\AA{$\; \buildrel \circ \over {\rm A}$}

\def\gtsim{\;\lower.6ex\hbox{$\sim$}\kern-6.7pt\raise.4ex\hbox{$>$}\;}
\def\ltsim{\;\lower.6ex\hbox{$\sim$}\kern-6.9pt\raise.4ex\hbox{$<$}\;}

\def\bmv{\hbox{\it B--V\/}}

\def\bmi{\hbox{\it B--I\/}}

\shorttitle{On a new method to estimate distance, reddening and metallicity of RR Lyrae stars}
\shortauthors{Bono et~al.}

\begin{document}

\title{On a new method to estimate distance, reddening and metallicity of 
RR Lyrae stars using optical/near-infrared ($B$,$V$,$I$,$J$,$H$,$K$) 
mean magnitudes: $\omega$ Centauri as a first test case} 

\author{
G.~Bono\altaffilmark{1,2},
G.~Iannicola\altaffilmark{2},
V.F.~Braga\altaffilmark{3,4},
I.~Ferraro\altaffilmark{2},
P.B.~Stetson\altaffilmark{5},
D. Magurno\altaffilmark{1},
N.~Matsunaga\altaffilmark{6},
R.L.~Beaton\altaffilmark{7},
R.~Buonanno\altaffilmark{8},
B. Chaboyer\altaffilmark{9},
M. Dall'Ora\altaffilmark{10},
M.~Fabrizio\altaffilmark{2,11},
G. Fiorentino\altaffilmark{12},
W.L.~Freedman\altaffilmark{13,7},
C.K.~Gilligan\altaffilmark{9},
B.F.~Madore\altaffilmark{7},
M.~Marconi\altaffilmark{10},
M.~Marengo\altaffilmark{14},
S.~Marinoni\altaffilmark{2,11},
P.~Marrese\altaffilmark{2,11},
C.E.~Martinez-Vazquez\altaffilmark{15},
M.~Mateo\altaffilmark{16},
M.~Monelli\altaffilmark{17,18},
J.R.~Neeley\altaffilmark{19},
M.~Nonino\altaffilmark{20},
C.~Sneden\altaffilmark{21},
F.~Thevenin\altaffilmark{22},
E.~Valenti\altaffilmark{23},
A.R.~Walker\altaffilmark{15}}
\altaffiltext{1}{Department of Physics, Universit\`a di Roma Tor Vergata, via della Ricerca Scientifica 1, 00133 Roma, Italy}
\altaffiltext{2}{INAF-Osservatorio Astronomico di Roma, via Frascati 33, 00040 Monte Porzio Catone, Italy}
\altaffiltext{3}{Instituto Milenio de Astrofisica, Santiago, Chile}
\altaffiltext{4}{Departamento de Fisica, Facultad de Ciencias Exactas, Universidad Andres Bello, Fernandez Concha 700, Las Condes, Santiago, Chile}
\altaffiltext{5}{NRC-Herzberg, Dominion Astrophysical Observatory, 5071 West Saanich Road, Victoria BC V9E 2E7, Canada}
\altaffiltext{6}{Kiso Observatory, Institute of Astronomy, School of Science, The University of Tokyo, 
10762-30, Mitake, Kiso-machi, Kiso-gun, 3 Nagano 97-0101, Japan}
\altaffiltext{7}{The Observatories of the Carnegie Institution for Science, 813 Santa Barbara St., Pasadena, CA 91101, USA}
\altaffiltext{8}{INAF-Osservatorio Astronomico d'Abruzzo, Via Mentore Maggini snc, Loc. Collurania, 64100 Teramo, Italy}
\altaffiltext{9}{Department of Physics and Astronomy, Dartmouth College, Hanover, NH 03755, USA}
\altaffiltext{10}{INAF-Osservatorio Astronomico di Capodimonte, Salita Moiariello 16, 80131 Napoli, Italy}
\altaffiltext{11}{SSDC, via del Politecnico snc, 00133 Roma, Italy}
\altaffiltext{12}{INAF-Osservatorio di Astrofisica e Scienza dello Spazio di Bologna, Via Piero Gobetti 93/3, 40129, Bologna, Italy}
\altaffiltext{13}{Department of Astronomy and Astrophysics, University of Chicago, 5640 S. Ellis Ave, Chicago, IL 60637, USA}
\altaffiltext{14}{Department of Physics and Astronomy, Iowa State University, Ames, IA 50011, USA}
\altaffiltext{15}{Cerro Tololo Inter-American Observatory, National Optical Astronomy Observatory, Casilla 603, La Serena, Chile}
\altaffiltext{16}{Department of Astronomy, University of Michigan, 1085 S. University, Ann Arbor, MI 48109, USA}
\altaffiltext{17}{Instituto de Astrof\'isica de Canarias, Calle Via Lactea s/n, E38205 La Laguna, Tenerife, Spain}
\altaffiltext{18}{Departamento de Astrofisica, Universidad de La Laguna, Tenerife, Spain}
\altaffiltext{19}{Department of Physics, Florida Atlantic University, Boca Raton, FL 33431, USA}
\altaffiltext{20}{INAF-Osservatorio Astronomico di Trieste, Via G.B. Tiepolo, 11, 34143 Trieste, Italy}
\altaffiltext{21}{Department of Astronomy and McDonald Observatory, The University of Texas, Austin, TX 78712, USA}
\altaffiltext{22}{Universite de Nice Sophia-antipolis, CNRS, Observatoire de la Cote d'Azur, Laboratoire Lagrange, BP 4229, F-06304 Nice, France}
\altaffiltext{23}{European Southern Observatory, Karl-Schwarzschild-Str. 2, 85748 Garching bei Munchen, Germany}


\date{\centering Submitted \today\ / Received / Accepted }

\begin{abstract}
We developed a new approach to provide accurate estimates of metal content,
reddening and true distance modulus of RR Lyrae stars (RRLs). The method is
based on  homogeneous optical ($BVI$) and near-infrared ($JHK$) mean magnitudes
and on predicted period--luminosity--metallicity relations ($IJHK$) and
absolute mean magnitude--metallicity relations ($BV$). We obtained solutions
for three different RRL samples in $\omega$ Cen: first overtone (RRc,~90),
fundamental (RRab,~80) and global (RRc+RRab) in which the period of first
overtones were fundamentalized.
The metallicity distribution shows a well defined peak at [Fe/H]$\sim$--1.98
and a standard deviation of $\sigma$=0.54 dex. The spread is, as expected,
metal-poor ([Fe/H]$\le$--2.3) objects. The current metallicity distribution
is $\sim$0.3 dex more metal-poor than similar estimates for RRLs available
in the literature. The difference vanishes if the true distance modulus we
estimated is offset by --0.06/--0.07~mag in true distance modulus.
We also found a cluster true distance modulus of $\mu$=13.720$\pm$0.002$\pm$0.030~mag,
where the former error is the error on the mean and the latter is the standard
deviation. Moreover, we found a cluster reddening of
E($B-V$)=0.132$\pm$0.002$\pm$0.028~mag and spatial variations of the order of
a few arcmin across the body of the cluster. Both the true distance modulus
and the reddening are slightly larger than similar estimates available in the
literature, but the difference is within 1$\sigma$. The metallicity dependence
of distance diagnostics agree with theory and observations, but firm constraints
require accurate and homogeneous spectroscopic measurements.
\end{abstract}


\maketitle

\section{Introduction} \label{chapt_intro_omega}

The use of RR Lyrae and classical Cepheids as first rungs in the cosmic 
distance scale dates back to more than one century ago 
\citep{leavitt08,hubble25,shapley53,baade56}. Fundamental contributions 
on the diagnostics adopted to estimate individual distances have been 
provided over more than half a century by \cite{sandage68,tammann03}.  
The empirical scenario was complemented during the eighties with the use 
of near-infrared mean magnitudes by \cite{longmore86,welch83,madore87}.  

The theoretical framework after the seminal investigations by Cox, Christy, 
Iben and Castor lagged until a proper treatment for the convective 
transport \citep{stellingwerf82,stellingwerf82b} was included in the calculation of 
radial pulsation models. The use of the new radiative and molecular 
opacities (OP, OPAL) together with a more formal treatment of the free 
parameters adopted for dealing with eddy viscosity and artificial viscosity
\citep{bono94b} paved the way to detailed and homogeneous predictions 
for radial variables in the Cepheid instability strip  
\citep{kovacs1990,bono99a,feuchtinger1999,smolec2013}.
The key advantage of the latter approach compared with the classical one 
is the opportunity to constrain the modal stability and the pulsation 
amplitudes. Moreover and even more importantly, the use of a common environment 
between evolutionary and pulsation prescriptions provided for the first time 
the opportunity to constrain the metallicity dependence of the diagnostics 
adopted to determine individual distances.  

In this context cluster RR Lyrae have played a crucial role, since 
their progenitors typically share the same ages and the same chemical 
composition distributions.
Moreover, they can be adopted to determine both the zero-point and the slope 
of optical ($R,I$, \citealt{braga16}, hereinafter BR16) and Near-Infrared 
(NIR, $J$,$H$,$K$, \citealt{braga2018}, hereinafter BR18) Period-Luminosity-Metallicity 
(PLZ) relations. These are key advantages when compared with field variables, 
since accurate metal abundances are only available for $\sim$100 objects 
\citep[e.g. ][Fabrizio et al. 2018, in prepration]{for11,pancino15,sneden2017}.  
Among the clusters hosting a sizable sample of RRLs, we will focus our attention on 
$\omega$ Cen (NGC~5139) for the following reasons.

a) Sample size -- $\omega$ Cen includes $\approx$200 RRLs that are almost 
equally split between fundamental and first overtone pulsators 
\citep[][BR18]{navarrete17}. 

b) Spread in iron abundance -- The current evidence indicates that RRLs 
in $\omega$ Cen cover a range in metallicity of at least one dex:
$-2.2\lesssim$[Fe/H]$\lesssim -1$, \citep{sollima06a}; 
$-2.4\lesssim$[Fe/H]$\lesssim -0.8$, \citep{rey2000}. 

c) Mean Magnitudes -- Our group provided a complete census of RRLs 
in $\omega$ Cen (BR16,BR18), this means new and homogeneous optical and NIR 
mean magnitudes together with the characterization of individual 
variables (luminosity amplitudes, periods, non-linear phenomena 
[mixed--mode, Blazhko]).  

d) Reddening variation -- There is mounting evidence that the 
cluster is affected by differential reddening \citep{calamida05}.  
 
These are the reasons why the quoted optical/NIR data have already 
been adopted to estimate the cluster distance 
(BR16,BR18) and the metallicity distribution of RR Lyrae (BR16).  
To overcome uncertainties on individual metal abundances and 
extinctions, BR16 evaluated the cluster distance using 
$V$,$B-V$ and $V$,$B-I$ Period-Wesenheit (PW) relations. These relations 
are, by construction, independent of uncertainties affecting 
cluster reddening and by the possible presence of differential 
reddening \citep{marconi15}. Moreover, their dependence on metallicity 
is small (<0.1 mag/dex) and they can be considered as metal-independent 
PW relations.
A similar approach was adopted by BR18, but the cluster distance 
was evaluated using NIR ($JHK$) mean magnitudes which are one order 
of magnitude less affected by uncertainties on individual reddenings
when compared with optical mean magnitudes. Moreover, they used metal 
abundances available in the literature based either on spectroscopy 
\citep{sollima06a} or on photometric indices \citep[][BR16]{rey2000}.  

To overcome some of the quoted limitations we introduce a novel 
approach based on six ($B$,$V$,$I$,$J$,$H$,$K$) mean magnitudes 
to provide distance, reddening, and metal abundance estimates of 
individual RRLs. The structure of the paper is as follows. 
In Section 2 we introduce the empirical and the theoretical frameworks 
on which this investigation relies. In Section 3 we discuss in detail 
the new approach we developed to provide homogeneous estimates of 
metal content, true distance modulus and reddening of individual 
RRLs. In particular, we discuss the different steps and assumptions 
we followed in the first and in second iteration of the method. 
Section 4 deals with the tests we performed to validate the new 
approach. In Section 5 we discuss the comparison with similar estimates 
available in the literature. In particular, the comparison with 
spectroscopic and spectro-photometric metallicity distributions is 
discussed in Section 5.1, while Section 5.2 deals with the reddening 
distribution and Section 5.3 with the true distance modulus distribution. 
Finally, Section 6 gives a summary of the current results together with 
a few remarks concerning the future developments of this project. 
 

\section{Empirical and theoretical framework}\label{chapt_obs_nir}

\begin{figure*}[!htbp]
\centering
\includegraphics[width=11cm, height=12cm]{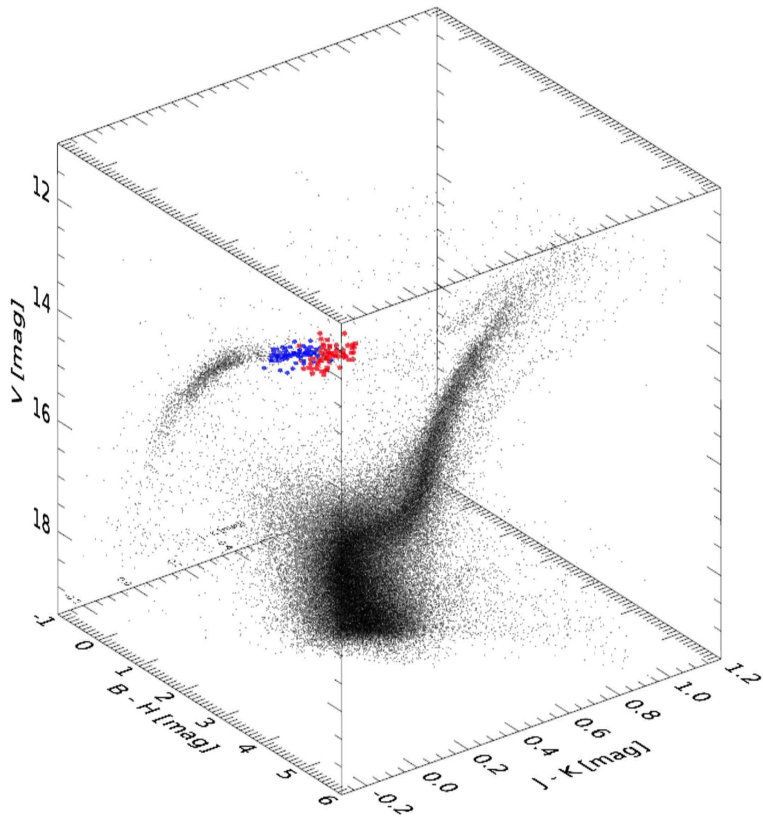}
\caption{
3D apparent Color-Magnitude Diagram---$V$,$B-H$,$J-K$---of $\omega$ Cen based on optical 
and NIR photometry provided by BR16 and BR18. Red and blue circles display the position 
in the instability strip of both fundamental (RRab) and first overtone (RRc) RRLs. 
The error on the mean magnitude is typically smaller than the symbol size. 
Black dots display $\omega$ Cen stars selected according to radial distance 
(5$\le$ r $\le$ 15 arcmin) and number of measurements (dozen per band). Only stars 
brighter than V=19.5 mag were plotted.  
}
\label{fig:omegacen_1}
\end{figure*}

The optical--NIR data adopted in the current investigations are shown in Fig.~\ref{fig:omegacen_1}. 
The 3D ($V$,$B-H$,$J-K$) Color-Magnitude diagram (CMD) displays a small fraction of 
cluster stars (black dots) and the position of the entire sample of 196 RRLs. The 
blue circles display first overtones (RRc), while the red circles the 
fundamentals (RRab). The reader interested in the complete census of RRLs concerning 
candidate Blazhko RRLs and the candidate mixed mode RRL is referred to BR16.

The Period-Luminosity (PL) and the Period-Wesenheit relations adopted in this paper 
rely on the grid of nonlinear, convective pulsation models computed by 
\cite{marconi15}. The main difference is that the same models were transformed 
into the observational plane using the very same $JHK$ passbands adopted by the 2MASS 
NIR photometric system\footnote{Note that throughout the paper we are using $K$ 
instead of the 2MASS $K_s$-band.}. This means that we did not use the transformations provided 
by \cite{campbell10} to move from the Bessell \& Brett photometric system into the 
2MASS system. The new NIR distance diagnostics will be provided in a forthcoming 
paper (Marconi et al. 2018, in preparation). 

It is quite well known that RRLs in the bluer optical ($BV$) bands obey to 
a mean magnitude--metallicity relation, while the PL relation becomes more evident 
for wavelengths longer than the $R$-band \citep{bono03c,catelan04,marconi15,braga15}. 
To fully exploit the optical data set we also derived new $B$ and $V$ mean 
magnitude--metallicity relations. They are based on the same set of RRL models 
by \cite{marconi15} and will be provided in Marconi et al. (2018, in preparation).

\section{REDIME: a new approach to estimate REddening DIstance and MEtallicity of RRLs}\label{chapt_rrl}

\begin{figure*}[!htbp]
\centering
\includegraphics[width=14cm, height=10cm]{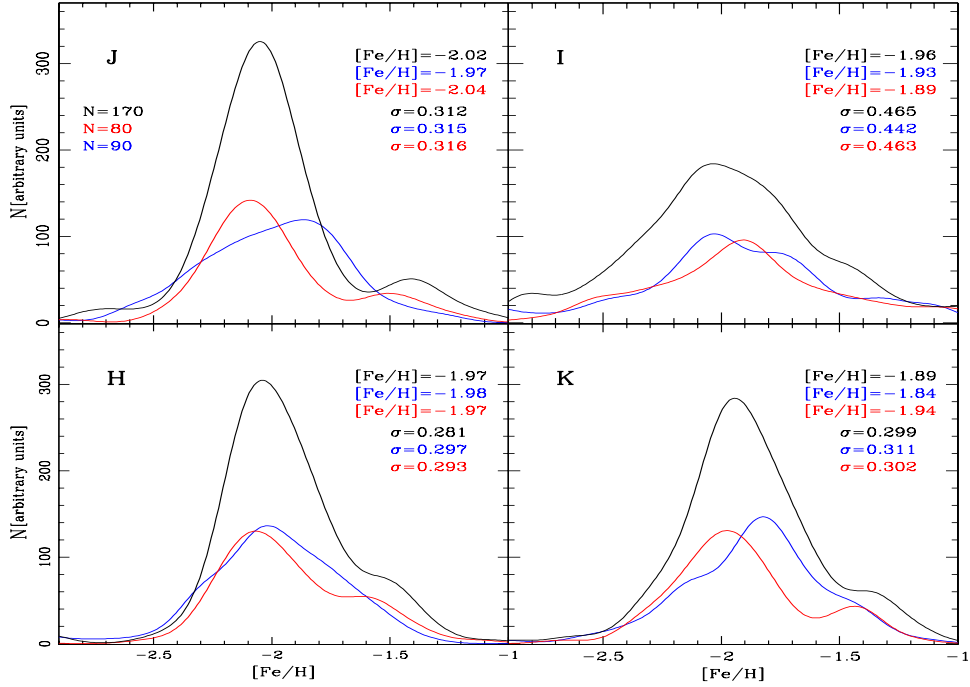}
\caption{
Metallicity distribution of RRLs in $\omega$ Cen using optical (I) and NIR 
($J$,$H$,$K$) Period-Luminosity Metallicity relations. To estimate the metal 
abundances we adopted a fixed true distance modulus ($\mu$=13.698 mag, BR18) 
and a mean cluster reddening \citep[E($B-V$)=0.11 mag][]{calamida05}. 
The metallicity distributions were smoothed using a Gaussian kernel 
with unit weight and $\sigma$ equal to the error of the individual estimates.    
The red and the blue lines display the metallicity distribution for 
RRab and RRc variables, while the black line the metallicity 
distribution of the global sample. The median values and the standard 
deviation of the metallicity distributions are also labeled.  
}
\label{fig:omegacen_2}
\end{figure*}

We developed a new approach to estimate REddening, DIstance modulus and MEtallicity
(REDIME) of field and cluster RRLs. REDIME is only based on optical and 
NIR measurements, but it can be easily extended to near UV and mid-infrared bands.   
The RRLs hosted by $\omega$ Cen are roughly 200 and our group collected during the 
last 15 years sizable samples of optical ($UBVRI$) and NIR ($JHK$) time series
(BR16,BR18).  
In spite of this unprecedented observational effort there are still RRLs for which 
either the optical and/or the NIR mean magnitudes are not very accurate.  These 
objects are typically located in the outskirts of the cluster. 
%
%
To avoid possible systematics in applying REDIME we restricted the sample
to the RRLs for which we have accurate optical and NIR mean magnitudes. This
means variables with a good coverage of the light curve with at least a ten
phase points per band. We ended up with a sample of 170 RRLs, listed in
Table 1. Among them, 90 are RRc and 80 are RRab. Note that the sample
also includes 26 candidate Blazhko RRLs. There is only one candidate
mixed mode RRL in $\omega$ Cen (Braga et al. 2018) and it was not included
in the current analysis. The properties of this interesting RRL variable
will be addressed in a separate paper (Braga et al. 2018, in preparation). 
The REDIME algorithm relies on two iterations. The former one is aimed at providing 
an initial homogeneous estimate of the metallicity, true distance modulus and reddening 
estimates. The latter to further improve both the precision and the accuracy of the 
initial guess. The individual steps performed to approach the final solution are 
summarized in the following. The reader interested in a more detailed description of 
the algorithm is referred to the flow chart presented in the Appendix to this paper.\\ 

{\bf First Iteration}--\\  

{\bf a)}-- We provided a new estimate of the cluster distance using the predicted 
($V$,$B-I$) PW relation. We adopted this relation, since it is independent of 
reddening uncertainties by construction. Moreover, theory and observations indicate 
that the metallicity dependence is negligible, and indeed the coefficient 
of the metallicity term is smaller than 0.1 dex. Furthermore, the standard 
deviation of this PW relation is smaller than the other optical ($V$,$B-V$) 
PW relations with a small coefficient of the metallicity term \citep{marconi15}.

{\bf b)}-- We provided a new estimate of the individual metallicities by inverting, 
using the new mean distance modulus and the mean cluster reddening (E($B-V$)=0.11 mag 
\citealt{calamida05}), four different optical/NIR ($I$,$J$,$H$,$K$) PLZ relations. 
The same approach, but only based on the $I$-band, was already adopted in the 
literature to estimate the RRL metallicity distribution in $\omega$ Cen (BR16) 
and in nearby dwarf galaxies \citep{martinezvazquez16a}. The inversion of the 
PLZ relations is straightforward and relies on the following equation: 

%
\begin{equation}\label{eq:invertedplz}
\mathrm{[Fe/H]}=\dfrac{M_X - b_X\log{P} - a_X}{c_X}
\end{equation}

where X is for the photometric band ($I$,$J$,$H$,$K$) while the constants $a_X$, $b_X$ 
and $c_X$ are the zero-point, the slope and the metallicity coefficient of the predicted 
PLZ relations ($M_X = a_X+b_X\log{P}+c_X\mathrm{[Fe/H]}$). The coefficients $a_X$, $b_X$ and 
$c_X$ are given by \citet{marconi15}  and by Marconi et al. (2018, in preparation).

Fig.~\ref{fig:omegacen_2} shows in anticlockwise direction the metallicity distributions based on 
inversion of the $I$,$J$,$H$,$K$ PLZ relations. The red and the blue lines display the metallicity 
distribution for RRab and RRc variables, while the black one for the global sample. In 
the global sample the period of RRc variables were fundamentalized, i.e.  
$\log P_F = \log P_{FO} + 0.127$ \citep{coppola15}. The metallicity distributions 
were smoothed using a Gaussian kernel with unit weight and $\sigma$ equal to the 
uncertainty on the metallicity. The median of the three different metallicity 
distributions together with their standard deviations are labeled.  

\begin{figure*}[!htbp]
\centering
\includegraphics[width=14cm, height=10cm]{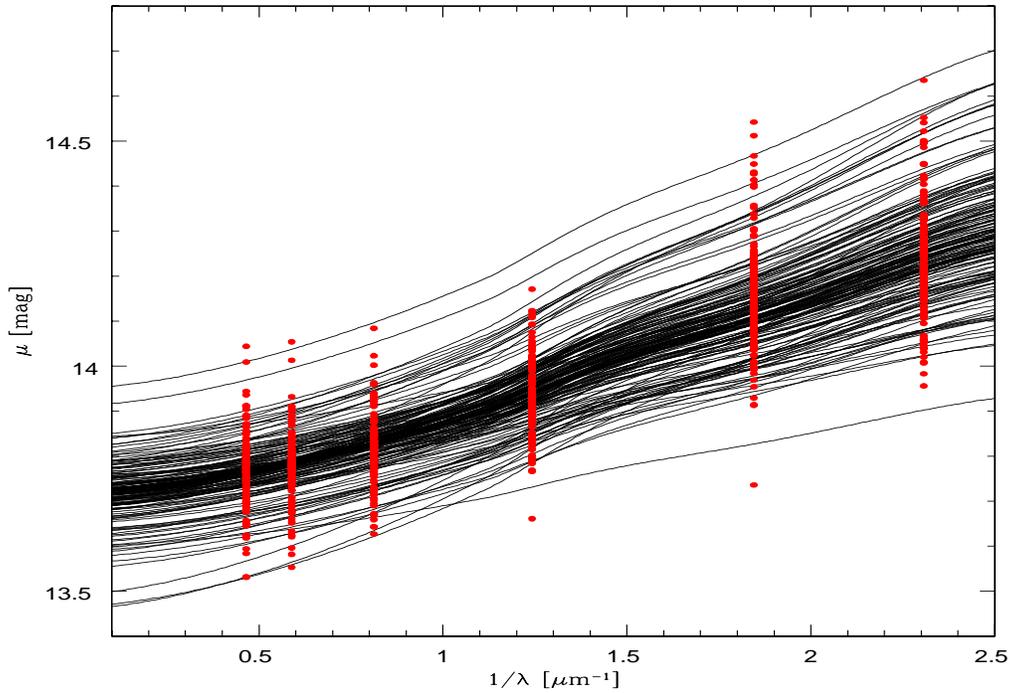}
\caption{
True distance modulus versus the inverse of the central wavelength 
of the adopted photometric bands. The solid lines display the individual nonlinear 
fits to the six different apparent distance moduli. From right to left the red dots 
display the apparent moduli based on optical ($B$,$V$,$I$) and on NIR ($J$,$H$,$K$) mean magnitudes.   
}
\label{fig:omegacen_3}
\end{figure*}

Data plotted in this figure show several interesting features. 

i) The standard deviations of the metallicity 
distributions based on NIR PLZ relations are systematically smaller than those 
based on the $I$-band PLZ relation. The difference is mainly caused by the increase 
in the slope and by the decrease in the intrinsic dispersion of the PLZ relations 
when moving from shorter to longer wavelength. Marginal variations in the mean 
cluster reddening might also contribute in explaining the minimal difference 
among optical ($I$) and NIR ($J$,$H$,$K$) estimates. 

ii) The metallicity distributions for RRc and RRab agree quite well, 
at fixed photometric band, with each other. Thus suggesting similar 
metallicity distributions.    

iii) The metallicity distributions show several secondary bumps, suggestive 
of a multimodal distribution. However, the position and the fraction of stars 
included in these secondary features changes among the different bands. In spite 
of these variations there is evidence of a shoulder in the more metal-rich regime 
for [Fe/H]$\ge$--1.5. A closer inspection into the metallicity distribution,  
based on different assumptions concerning the smoothing, indicates that the 
current estimates display a well defined secondary peak for [Fe/H]$\sim$--1.5.  
This finding further supports the evidence that the RRL metallicity distribution 
in $\omega$ Cen is at least bi-modal.    
In passing we also note that the metal-poor tail ([Fe/H]$\le$--2.5) 
vanishes when moving from the $I$-band to the $J$,$H$,$K$ bands. Thus suggesting 
that it might be affected by small reddening variations and/or uncertainties 
in the mean magnitudes.

\begin{figure*}[!htbp]
\centering
\includegraphics[width=14cm, height=10cm]{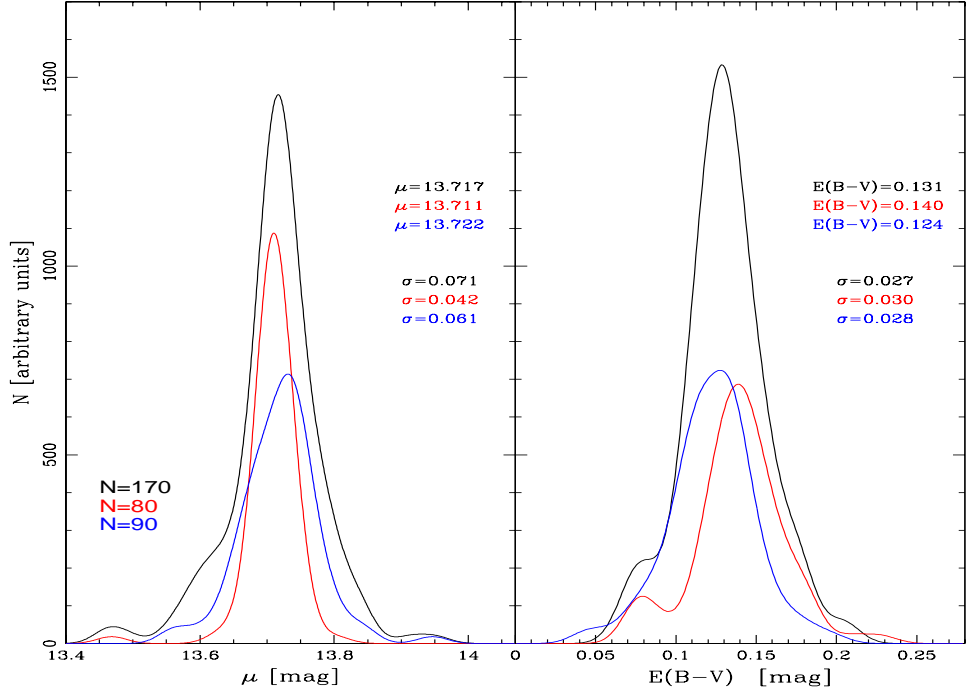}
\caption{
Left--True distance modulus distribution of $\omega$ Cen RR Lyrae stars. The 
individual distances were estimated using two mean magnitude--metallicity relations 
($M_B$,$M_V$) and four Period-Luminosity-Metallicity relations ($I$,$J$,$H$,$K$). The 
nonlinear fit was performed using the analytical reddening law provided by 
\cite{cardelli89} and extinction coefficients provided by \citep{stetson14a}. 
The individual metallicity evaluations adopted 
to estimate the distances are discussed in Section~\ref{chapt_rrl}. The red  
and the blue lines display the distance distribution for RRab and RRc, while 
the black line the global solution. The number of RRLs adopted for the three 
different solutions are labeled in the bottom left corner together with 
the median and the standard deviations.  
The distance distributions were smoothed using the same approach adopted 
to smooth the metallicity distributions. 
Right--Same as the left, but for the reddening. 
The individual reddening estimates were simultaneously estimated with the 
true distance modulus performing the nonlinear fit with the Cardelli's 
reddening law. The median and the standard deviations of the reddening 
distributions are also labeled.  
}
\label{fig:omegacen_4}
\end{figure*}
 
%
{\bf c)}-- The individual metallicities were estimated as the mean of the 
values based on the global relations in the three NIR ($J$,$H$,$K$) bands.  
The uncertainty was assumed equal to the mean standard deviation and it 
is typically smaller than 0.3 dex. Note that in estimating the error on 
the metallicity estimates we neglected the standard deviations of the 
predicted PLZ relations, since they are of the order of a few hundredths 
of a magnitude in the NIR bands \citep[see Table~6 in][]{marconi15}.   
On the basis of the new individual mean metal abundances, we applied 
the same approach adopted by \cite{inno2016} to estimate simultaneously 
the reddening and the true distance modulus. In particular, we seek to 
optimize the value of:

\begin{equation}
\chi ^2 = \sum_{j=1}^{6} \left( \frac{{\mu_j} - (\mu + \xi_j\,E(B-V))}{\sigma_j} \right)^2
\end{equation}

where the sum runs over the four PLZ relations ($I$,$J$,$H$,$K$) and the two MZ relations ($B$,$V$),
$\mu_j$ are the apparent distance moduli, $m_j-M_j$, in which the absolute 
magnitudes are based on pulsation predictions by \cite{marconi15}, $\mu$ is the true distance 
modulus, $\xi_j$ are the extinction coefficients according to the reddening law 
provided by \cite{cardelli89} with a constant total-to-selective absorption ratio 
($R_V$=3.1), and $\sigma_j$ are the standard deviations taking account for uncertainties 
on apparent mean magnitudes, on absolute magnitudes (standard deviations of both PLZ and MZ relations) 
and on the extinction coefficients (Inno et al. 2016). 
The two free parameters are $\mu$ and $E(B-V)$.    

Fig.~\ref{fig:omegacen_3} shows the individual nonlinear fits over the the six 
optical/NIR measurements in the true distance modulus versus the inverse of the 
central wavelength. 
This means that for each variable in our sample we performed in this plane a 
nonlinear fit (black line), based on the reddening law by \cite{cardelli89}, 
over its six optical/NIR mean RRL magnitudes.  
Data plotted in this figure show that the true distance modulus 
($\lambda \longrightarrow \infty $) is mainly constrained by longer 
wavelength $J$-, $H$-, and in particular, $K$-band mean magnitudes. The slope of the 
nonlinear fit, and in turn, the reddening is mainly constrained by shorter 
wavelength $B$-, $V$- and $I$-band mean magnitudes. 
This "star-by-star" multi-wavelength, reddening and true distance modulus plot 
was first introduced by \citet{rich2014} in 
their parallel study of Cepheids in NGC 6822 (see their Figures 5 and 6). 
The difference in the dispersion when moving from optical to NIR measurements 
is due to intrinsic and extrinsic effects.  

{\em Intrinsic} The dispersion in magnitude, at fixed stellar parameters 
(stellar mass, luminosity, chemical composition), decreases when moving 
from the optical to the NIR bands. This is because NIR bands are less 
prone to uncertainties caused by evolutionary effects (off-ZAHB evolution).  

{\em Extrinsic} Optical light curves are more prone to uncertainties caused 
by a non-optimal coverage of the light curve, because the luminosity amplitude 
steadily increases when moving from the $K$ to the $B$ band. 
Moreover, differential changes in the mean cluster reddening manifest 
themselves to a larger degree at shorter wavelengths.

The left panel of Fig.~\ref{fig:omegacen_4} shows the true distance modulus distribution for the
three different samples: RRc, RRab and global. The distance distributions 
agree quite well with each other, and indeed, the difference in the 
median value is of the order of 1\%. They are also quite symmetric and the 
standard deviations also attain similar values.      
On the other hand, the reddening distributions plotted on the right panel of the 
same figure suggest that the RRab variables (red line) seem to have reddenings 
larger than RRc (blue line) variables. However, the difference is of the order of 
0.5$\sigma$. As expected the color excess of the global sample attains reddening 
values that are intermediate between RRab and RRc variables. 

The mean iron abundance, the true distance modulus and the reddening after the 
first iteration for the three different samples are listed in columns 2,3 and 4 
of Table~2 together with their means. \\

{\bf Second Iteration}--\\ 

{\bf d)}-- On the basis of the new median reddening and true distance modulus 
(global sample) we provided a new estimate of the individual metal abundances by 
inverting once again the $I$,$J$,$H$,$K$ PLZ relations. The median and the $\sigma$ of the 
new metallicity distributions agree quite well with the metallicity distributions 
we obtained in the first iteration of REDIME. The agreement applies not only to the 
global sample, but also to the RRc and to the RRab sample. Indeed, the difference 
is on average smaller than 0.1 dex. Thus suggesting that the solutions are quite stable.      

\begin{figure*}[!htbp]
\centering
\includegraphics[width=14cm, height=10cm]{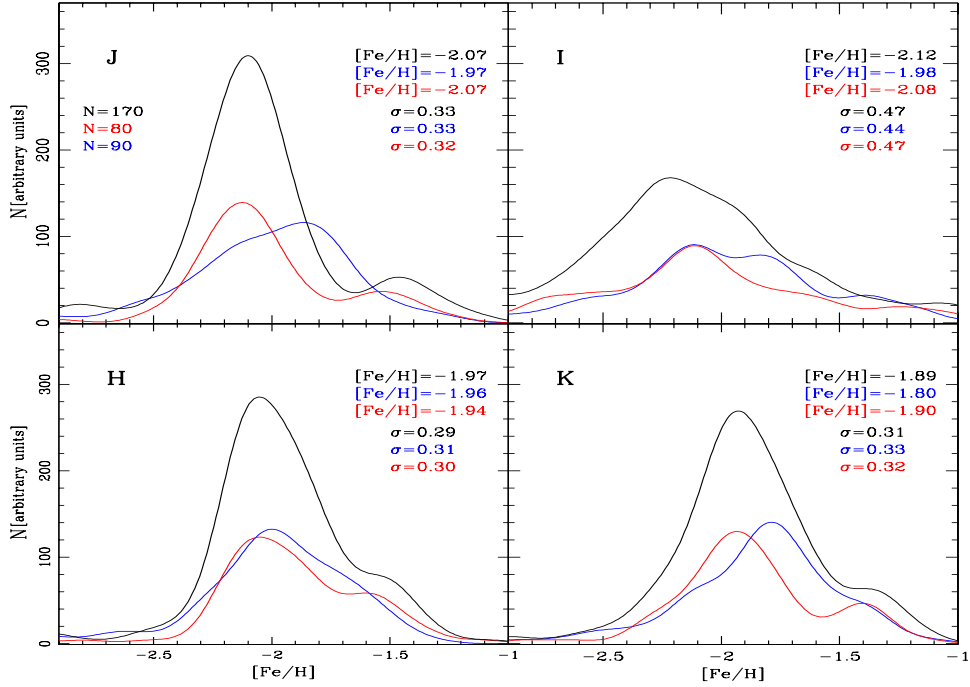}
\caption{
Same as Fig.~\ref{fig:omegacen_2}, but the metallicity estimates are based on individual 
reddening determinations and true distance moduli of RRLs obtained in 
the first iteration of REDIME. 
}
\label{fig:omegacen_5}
\end{figure*}

The metallicity distributions plotted in Fig.~\ref{fig:omegacen_5} indicate that metallicity 
estimates based on NIR diagnostics have standard deviations that are 0.1 dex 
smaller when compared with the $I$-band. Moreover, they are also quite homogeneous, 
and indeed the difference in standard deviations is at most a few hundredths 
of a dex. This is the reason why we performed a mean of the 
NIR bands and we found $<$[Fe/H]$>$=--1.98$\pm$0.05 and a standard deviation 
$\sigma$=0.54 dex. The first error is 
the error on the mean and it is quite small due to the sample size. The second 
error is the standard deviation of the metallicity 
distribution and it is mainly caused by the intrinsic spread in metal 
abundance of stellar populations in $\omega$ Cen 
\citep{hughes2000,johnson_e_pilachowski2010}. 
The solutions for the three different samples (RRc, RRab, global) are 
given in column 5 of Table~2.   

\begin{figure*}[!htbp]
\centering
\includegraphics[width=14cm, height=9cm]{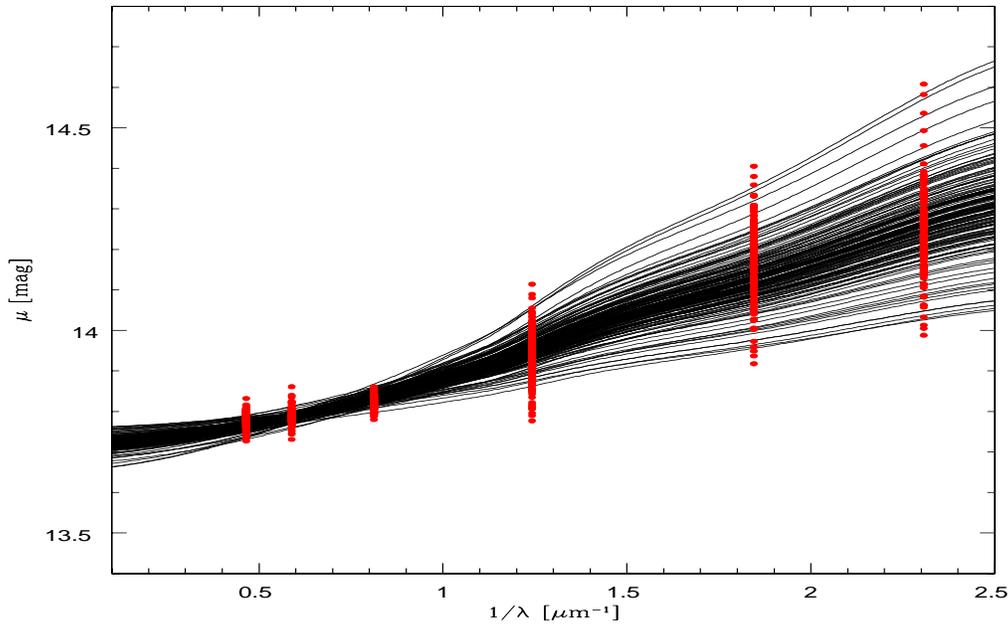}
\caption{
Same as Fig.~\ref{fig:omegacen_3}, but for solutions obtained in the 2nd iteration of REDIME. 
}
\label{fig:omegacen_6}
\end{figure*}

{\bf e)}-- The new individual mean metal abundances were used to perform new nonlinear 
fits (equation 2) of the six mean magnitudes. The individual fits plotted 
in Fig.~\ref{fig:omegacen_6} display the 
significant improvement in the distance modulus and reddening solution when moving 
from the first to the second iteration. The difference in lower and upper envelope 
in true distance moduli decreases from roughly $\sim$ 0.5 to less than 
0.1 mag in the NIR regime and from $\sim$0.8 to $\sim$0.6 mag in the optical regime.  
This evidence is further supporting the 
improvement on the individual mean metallicities. Note that the metallicity estimates
are only based on the inversion of $J$,$H$,$K$-band PLZ relations, while the simultaneous 
solution for distance and reddening also relies on two MZ relations ($B$,$V$) and on the 
$I$-band PLZ relation. 

\begin{figure*}[!htbp]
\centering
\includegraphics[width=14cm, height=9cm]{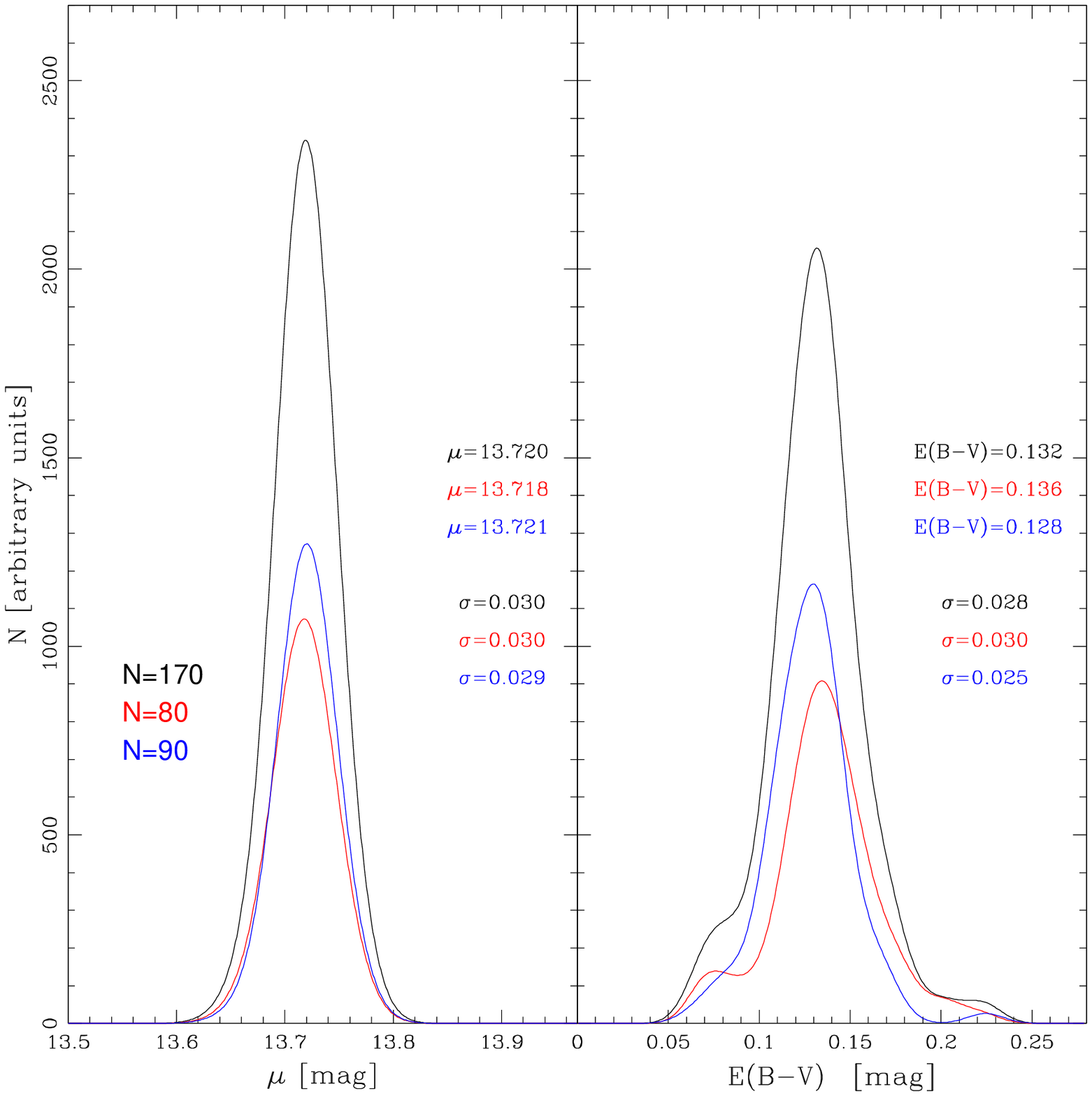}
\caption{
Same as Fig.~\ref{fig:omegacen_4}, but for true distance moduli and reddening estimates 
obtained on the 2nd iteration of REDIME. 
}
\label{fig:omegacen_7}
\end{figure*}

{\bf f)}-- The improvement between the first and the second iteration of REDIME 
becomes even more clear comparing the distribution of the true distance moduli 
plotted in the left panel of Fig.~\ref{fig:omegacen_4} 
and of Fig.~\ref{fig:omegacen_7}. The median cluster true 
distance moduli among the three different samples agree at the level of 1\%. 
The $\sigma$ of the global sample is a factor of two smaller when compared with 
the distribution obtained at the first iteration. 

The new reddening distributions plotted in the right panel of Fig.~\ref{fig:omegacen_7}
agree quite well with those based on the first iteration, further supporting the 
stability of the solution. We also performed a third iteration, but the results are, 
within the errors, identical to the second one.

\section{Internal consistency}\label{par:consistency}

\begin{figure*}[!htbp]
\centering
\includegraphics[width=14cm, height=10cm]{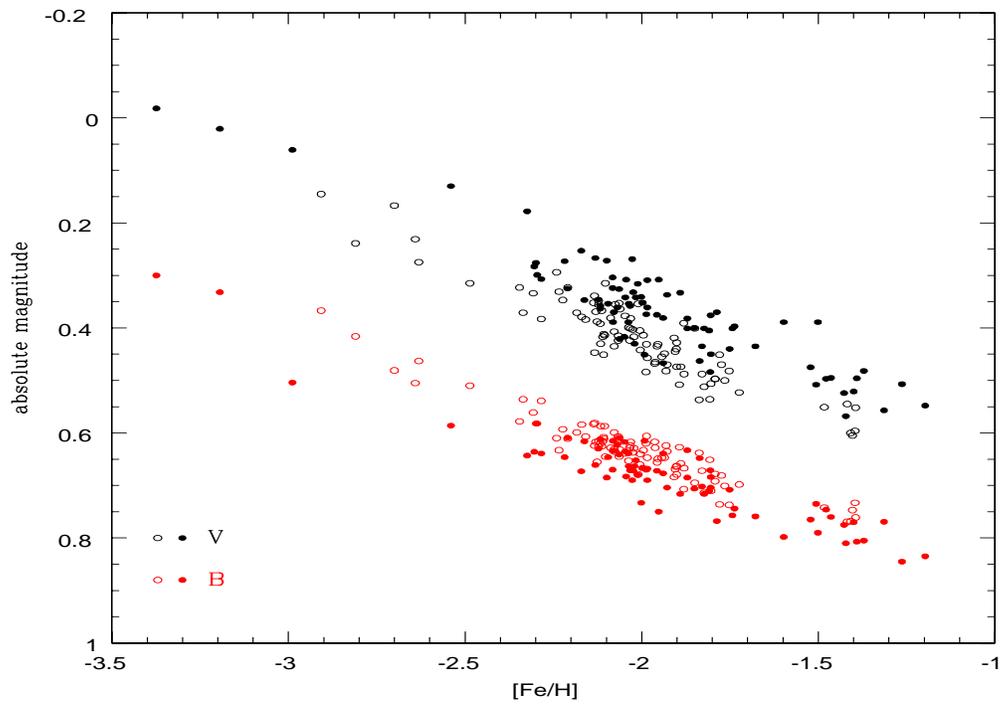}
\caption{
Absolute mean optical magnitude versus metallicity for $\omega$ Cen RRLs. Black and red 
dots display $B$- and $V$-band mean magnitudes, while filled and empty circles display 
fundamental (RRab) and first overtone (RRc) variables. The accuracy on the mean 
optical magnitudes is similar to the symbol size, on average better than one hundredth 
of a magnitude.  
}
\label{fig:omegacen_8}
\end{figure*}

To further constrain the internal consistency of 
REDIME, Fig.\ref{fig:omegacen_8} displays the 
distribution of the current sample of RRLs in the absolute mean magnitude-metallicity
plane. The RRLs display, as expected, a steady increase in the absolute mean magnitude 
as a function of the metal content. Data plotted in this figure 
also show that the spread in visual magnitudes is, at fixed metal content, is systematically 
larger than in the $B$-band.  
%

There is also evidence that RRc (empty symbols) variables in the $V$ band 
and for metal abundances ranging from [Fe/H]$\sim$--2.4 to [Fe/H]$\sim$--1.6 are, 
at fixed magnitude, systematically more metal-poor than RRab (filled symbols) 
variables. The trend is not very well defined in the more metal-rich ([Fe/H]>--1.6) 
and in the more metal-poor  ([Fe/H]>--2.4) regime, due to the paucity of objects.
Preliminary plain physical arguments based on the sensitivity of the HB morphology to 
metal content, might suggest that RRc variables, being systematically hotter than 
RRab variables, are more associated with more metal-poor stellar populations in 
$\omega$ Cen.  
The empirical scenario is far from being fully understood, and indeed, in the 
$B$ band, RRc and RRab variables display similar trends---when compared with 
the $V$ band---over the entire metallicity range. Thus suggesting a different 
sensitivity to the metal content when compared with the $V$ band.

\begin{figure*}[!htbp]
\centering
\includegraphics[width=14cm, height=9cm]{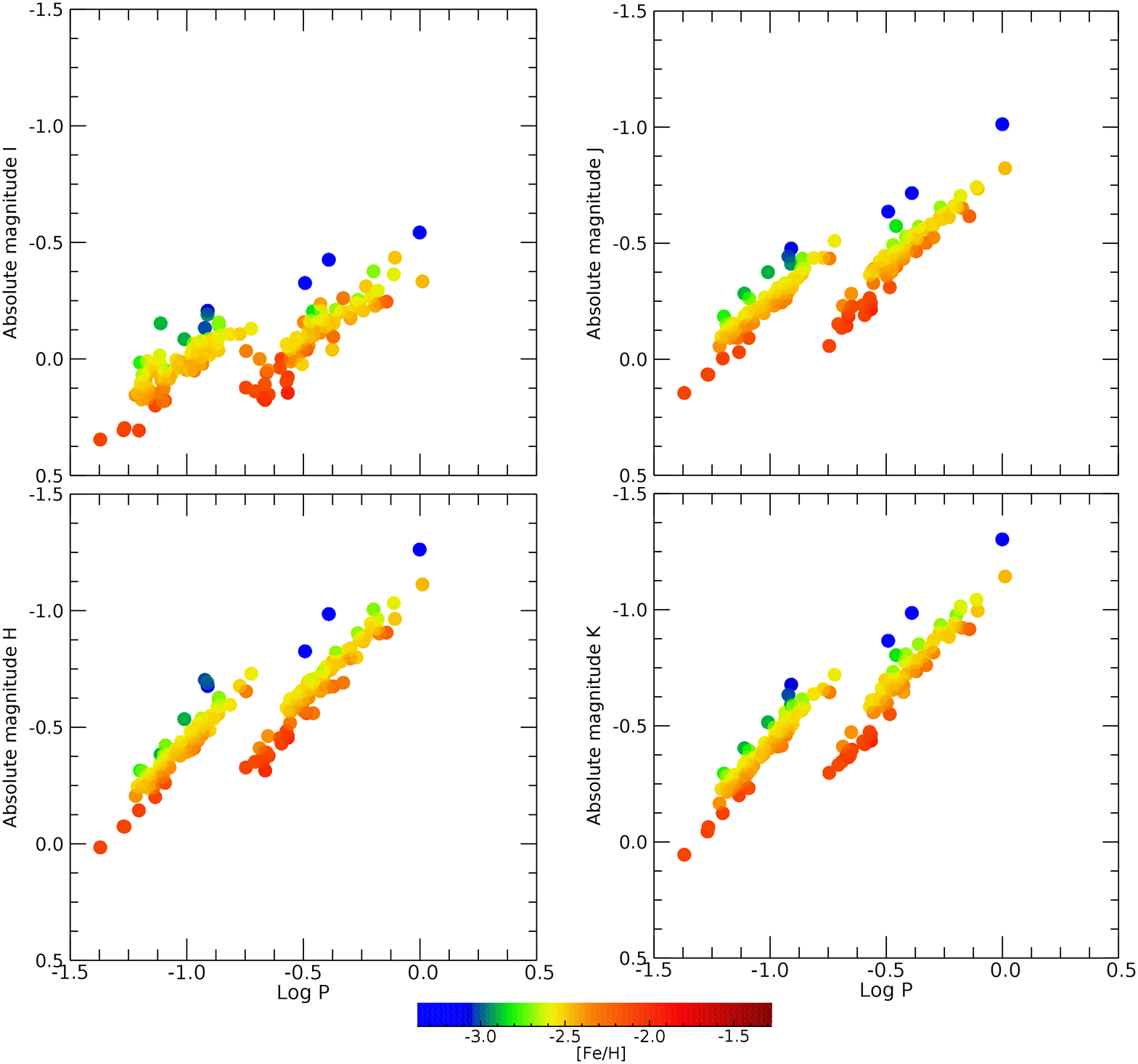}
\caption{
Period-Luminosity relations for $\omega$ Cen RRLs. 
The color of the individual RRLs is color coded according to their 
metallicity and the scale is displayed on the bottom of the figure. 
}
\label{fig:omegacen_9}
\end{figure*}

The internal agreement in absolute mean magnitude, reddening and metallicity 
estimates is soundly supported by the optical ($I$) and NIR ($J$,$H$,$K$) data plotted 
in the Period-Luminosity plane (see Fig.~\ref{fig:omegacen_9}). Note that the slope is 
well defined for both RRc and RRab variables and the spread in magnitude, at fixed 
period, is quite limited for both optical and NIR mean magnitudes.

\section{Comparison with literature values}\label{par:comparison}

\subsection{Metallicity distribution}\label{metallicity_one}
%
To validate the photometric metallicity estimates based on REDIME we 
performed a detailed comparison with similar abundance estimates available 
in the literature.  We selected the spectroscopic sample provided by 
\cite{sollima06a} and the photometric sample provided by \cite{rey2000}. 
To provide a homogeneous metallicity scale the iron abundances provided by 
S06 and R00 were rescaled to the cluster metallicity scale provided by 
Carretta et al. (2009). The iron abundances by S06 were rescaled by taking 
into account the difference in solar iron abundance in
number $\log\epsilon_{Fe}$=7.52 versus 7.54
\citep{gratton2003,carretta09}. The iron abundances by R00 were
transformed from the \citet{ZW84} metallicity scale into the
\citet{carretta09} metallicity scale using the linear relation
given in their \S~5. Moreover, the iron abundances based on the inversion
of the PLZ relations were rescaled from the solar iron number abundance
$\log\epsilon_{Fe}$=7.50 adopted in pulsation \citep{marconi15}
and in evolutionary models \citep{pietrinferni2006} to 7.54 of the 
Carretta et al. metallicity scale.

\begin{figure*}[!htbp]
\centering
\includegraphics[width=12cm, height=12cm]{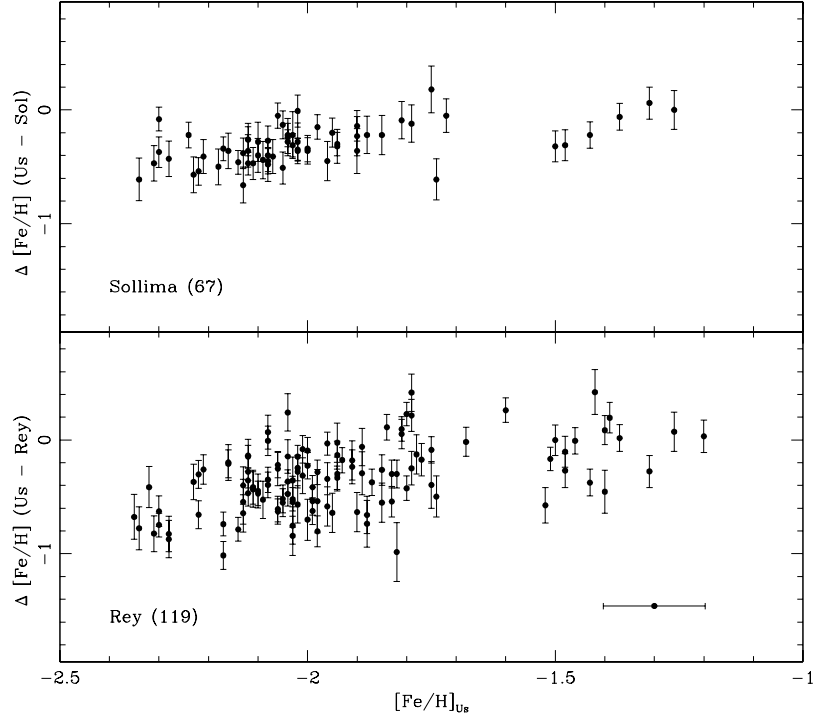}
\caption{
Top -- Comparison between the individual metal abundances based on REDIME and on 
spectroscopic measurements provided by \cite{sollima06a}. The vertical error bars 
display the error in quadrature of both spectroscopic and photometric errors. The 
error bar plotted in the bottom right corner shows the mean uncertainty on 
metallicity estimates based on REDIME. The number in parentheses show the 
number of objects in common.  
Bottom -- Same as the top, but the comparison is with the individual metal abundances 
based on the spectro-photometric estimates provided by \cite{rey2000}.  
}
\label{fig:omegacen_10}
\end{figure*}

The top panel of Fig.~\ref{fig:omegacen_10} shows the comparison between the
current metallicity estimates and the spectroscopic measurements
provided by \cite{sollima06a}. The comparison was performed for the 67 RRLs
in common in the two samples and we found that the difference is within
1$\sigma$. However, data plotted in this panel display that REDIME abundances
are, on average, 0.35 dex more metal-poor than the spectroscopic ones and
the difference is mainly in the zero-point.
The bottom panel of the same figure shows the comparison between REDIME
metallicities and the photometric estimates provided by \cite{rey2000}.
The comparison for the 119 RRLs in common shows the same trend already
found in the comparison with the spectroscopic sample. Indeed, the
difference is once again a difference in the zero-point ($\Delta$ [Fe/H]$\approx$0.40 dex).
Data plotted in Fig.~\ref{fig:omegacen_10} indicate a marginal evidence for a
possible systematic trend when moving from more metal-poor to more metal-rich RRLs.
However, the uncertainties on individual metallicities are still too large (see
error bars) to reach a firm conclusion.

\begin{figure*}[!htbp]
\centering
\includegraphics[width=14cm, height=9cm]{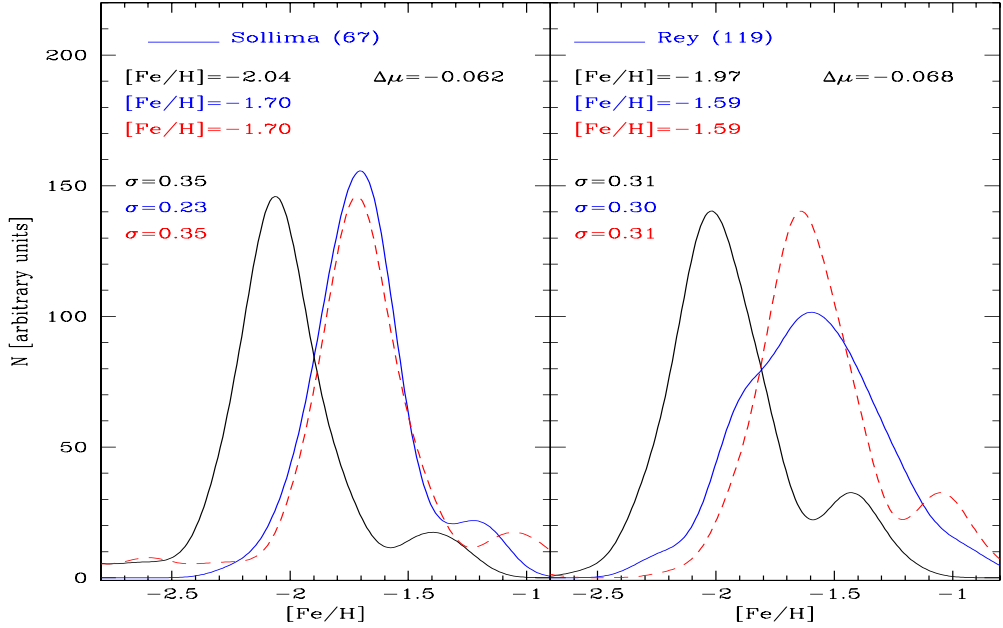}
\caption{
Left -- Comparison between the metallicity distribution based on REDIME (black line) 
and on spectroscopic measurements (blue line) for RRLs provided by \cite{sollima06a}. 
The dashed red line shows the metallicity distribution based on RRLs once we assume a 
true distance modulus that is 0.062 mag fainter than the true distance modulus based on 
REDIME. The peaks, the standard deviations of the different metallicity distributions 
and the number of objects in common are also labeled. 
Right -- Same as the left, but the comparison is with the metallicity distribution based 
on spectro-photometric measurements provided by \cite{rey2000}.  
}
\label{fig:omegacen_11}
\end{figure*}

To further constrain the precision of the current metallicity scale, the left
panel of Fig.~\ref{fig:omegacen_11} shows the comparison between the metallicity
distributions based on REDIME and on the spectroscopic measurements by
\cite{sollima06a}. Once again the comparison was restricted to the 67 RRLs in common
and we found that the two distributions agree within 1$\sigma$. The distribution
plotted in this panel display that the difference is, as expected, mainly a
difference in zero-point. Moreover, the standard deviation of the metallicity
distribution based on REDIME abundances is 0.1 dex larger than the spectroscopic
one. We have already mentioned in Section~\ref{par:consistency}
that REDIME is prone to possible systematics in the zero-point of the
adopted distance scale. Therefore, we performed a number of simulations to constrain
this effect and we found that a decrease of 0.062 mag in the true distance modulus
would provide a metallicity distribution (red dashed line) that agrees quite
well with the spectroscopic distribution (see labeled values).

%
We performed the same comparison with the spectro-photometric estimates provided
by \cite{rey2000}. The comparison shows the same trend already
found in the comparison with the spectroscopic sample. The difference is mainly in the
zero-point, but the standard deviations of the two metallicity distributions are
quite similar. We performed a number of simulations and we found that a decrease
of 0.068 mag in the true distance modulus would provide a metallicity distribution
(red dashed line) that agrees quite well with the photometric distribution
(see labeled values). 

The above findings indicate that we are facing two possible routes. 
a) The metallicity distribution based on REDIME is $\approx$0.35 dex systematically 
more metal-poor than spectroscopic and spectro-photometric metallicity distributions 
available in the literature. 
b) The current cluster true distance modulus is over-estimated by 0.062 and 0.068 mag
due to a systematic offset in the predicted zero-point of the RRL distance scale. 
Independent spectroscopic estimates covering a broader metallicity range 
\citep{chadid2017,sneden2017} are required to investigate whether 
the quoted difference is caused by uncertainties either on metallicities 
or on distance modulus estimates.   

\subsection{Reddening distribution}\label{reddening_one}

We also decided to compare the RRL reddenings based on REDIME with the reddening 
evaluations recently provided by Gaia DR2 \citep{gaia_dr2}. The reason 
why  we decided to use the reddening given in the general source catalog 
instead of the reddening provided for the RRLs is twofold. 

{\em i)} --- The number of $\omega$ Cen RRLs present in the Gaia catalog
for variable stars is quite limited (97 out of 198). Moreover, the light
curves and the pulsation parameters are not very accurate due to crowding
and phase coverage problems.

{\em ii)} --- To properly evaluate the reddening distribution across the 
body of the cluster we first selected in the Gaia source catalog the candidate 
cluster stars by using the new proper motion measurements. We plotted all the 
$\omega$ Cen sources within the truncation radius of $\omega$ Cen 
($r_t$=57.03 arcmin, \citealt{harris96} and updates). Candidate cluster stars 
were identified as a secondary maximum in proper motion plane with a centroid 
located at $\mu_{\alpha*}=-3.18$ mas yr$^{-1}$ and $\mu_{\delta}=-6.72$ mas yr$^{-1}$. 
Note that the current estimates agree quite well with the proper motion estimate 
($\mu_{\alpha*}=-3.1925\pm0.0022$ mas yr$^{-1}$, $\mu_{\delta}=-6.7445\pm0.0019$ mas yr$^{-1}$) 
provided by Gaia DR2 \citep{gaia_dr2}.
The stars brighter than G=16.5 mag located within 1.16 mas/yr of the centroid 
position were considered candidate cluster stars. We plotted this sample in a 
3D magnitude-color-color plot---$G$,$GBP$-$GRP$---and we selected the stars belonging 
to the $\omega$ Cen cluster sequences. We ended up with a sample of $\sim$3700 
stars and we found that the E($GBP$-$GRP$) is centered on 0.117 mag and the standard 
deviation is $\sigma$=0.088 mag. Note that in performing this fit we neglected 
the very high reddening tail  of the distribution (E($GBP$-$GRP$) > 0.3 mag).   

\begin{figure*}[!htbp]
\centering
\includegraphics[width=17cm, height=10cm]{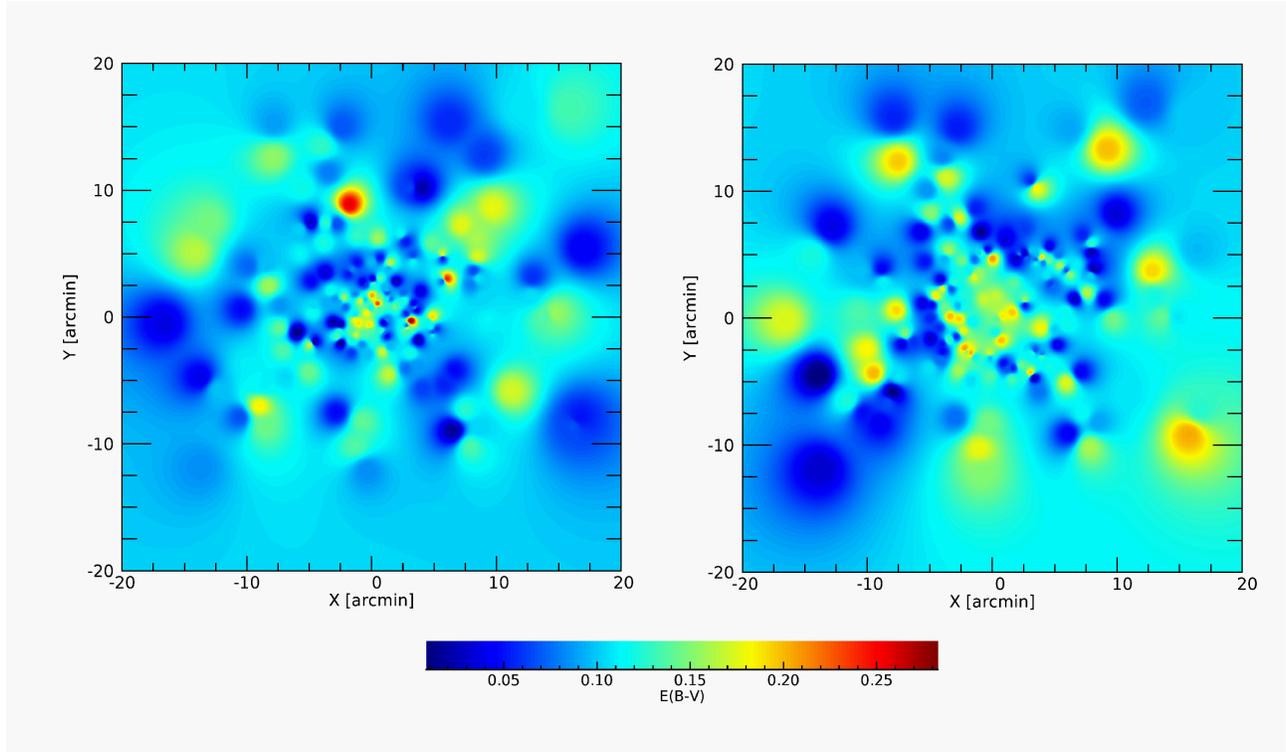}
\caption{
Left -- Reddening distribution across the body of $\omega$ Cen. The individual mean 
reddening estimates are based on REDIME and are color coded according to the 
reddening value (scale on the bottom). The reddening distribution was estimated by 
using the relative RRL distance (see text for more details).  
Right -- Same as the left, but the reddening distribution was estimated by using  
reddening estimates provided by Gaia DR2 \citep{gaia_dr2}. To avoid possible systematics in 
the reddening distribution, cluster stars were selected according to radial distance and 
proper motion.  Moreover, to compare two samples with similar spatial distributions 
for each RRL in our sample we only selected the closest static star.  
}
\label{fig:omegacen_12}
\end{figure*}

To compare the reddening distribution based on RRLs and on Gaia estimates, 
we transformed the E($B-V$) into E($GBP$-$GRP$) by using the Cardelli's empirical 
reddening law and the following extinction coefficients:  
$A_G/A_V$=0.840 mag, $A_{BP}/A_V$=1.086 mag, and $A_{RP}/A_V$=0.627 
mag\footnote{We adopted the new measurements of the Gaia passbands provided by
\citet{evans2018} and performed a polynomial fit to estimate the central 
wavelengths. We found $\lambda_c(G)\sim6420$ \AA, $\lambda_c(GBP)\sim5130$ \AA,  
$\lambda_c(GBP)\sim7800$ \AA,}. 
%
The reddening in the Gaia source catalog was estimated using the spectral
energy distribution of individual sources. The interstellar absorption in the
Gaia Source catalog was estimated using the three photometric bands
(G, G$_{BP}$. G$_{RP}$) and the parallax. The approach relies on the application
of a machine learning algorithm to a training data set including stars
characterized by low extinctions and for which the effective temperature
was typically spectroscopically estimated (see Section 2.1 in \citet{andrae2018}.
We found that the reddening distribution is centered on 0.115 mag, and the standard
deviation is $\sigma$=0.065 mag. The two independent reddening distributions are
in reasonable agreement, in particular, if we account the difference in sample
size (170 vs 3700), in spatial distribution and in the adopted photometric system.

To further investigate the reddening variation across the field of view, we
plotted the spatial distribution of RRLs investigated with REDIME on sky
(see the left panel of Fig.\ref{fig:omegacen_12}). The reddening is color
coded and the scale is displayed on the bottom.
To overcome the limitation of discrete sampling, the reddening map was
computed using a bi-dimensional grid with a bin of $\sim$2.5 arcsec. The reddening
of individual grid points was estimated averaging the reddening of RRLs in
the entire sample according to the radial distance between the grid point
and all the RRL in our sample. Closer is the RRL, larger is its contribution
to the mean reddening of the grid point. We followed this approach to estimate
the reddening map, because it naturally smooths the reddening distribution
in the cluster regions covered by RRLs.

The reddening map showed in the left panel of Fig.\ref{fig:omegacen_12}
indicates that the largest variations are across second and
third quadrant (X from --5 to --10 arcmin, Y from --5 to 12 arcmin) in which
the reddening changes from a few hundredths of a magnitude to a few tenths
of a magnitude (E($B-V$)$\sim$0.25 mag). These findings further support
the evidence that the extinction towards $\omega$ Cen changes on spatial
scales of the order of a few arcminutes or even less.
Moreover, there is no clear evidence of a radial extinction gradient in
the cluster region covered by the current sample. The lack of a gradient
either in reddening and/or in metal abundance should be cautiously treated,
since we are missing the RRLs located in the outskirt of the cluster
\citep{fernandeztrincado2015}.

To further investigate the reddening variation across the body of the cluster,
the right panel of the same figure shows the map based on reddening
estimates provided by Gaia DR2. Note that to properly compare the two samples
for each RRL we only selected the closest static star in the Gaia source
catalog. The two reddening maps display similar variations. There is no clear
evidence of an extinction gradient and the spatial variations of the reddening
quite similar to the map based on RRLs. Moreover, the mean reddening of the
170 static stars is 0.107 mag and its standard deviation is
0.055 mag. The difference with the estimates based on the entire sample
(3700 sources) is minimal concerning the mean reddening (0.107 vs 0.115),
but the standard deviation decreases from 0.065 to 0.055 mag. The new
standard deviation is still larger than the standard deviation based
on RRLs (0.28 mag). The difference is mainly caused by the
limited sample of stars in the Gaia source catalog located in the
innermost regions of the cluster due to the extreme crowding.

\subsection{True distance modulus distribution}\label{reddening_one}

\begin{figure*}[!htbp]
\centering
\includegraphics[width=14cm, height=10cm]{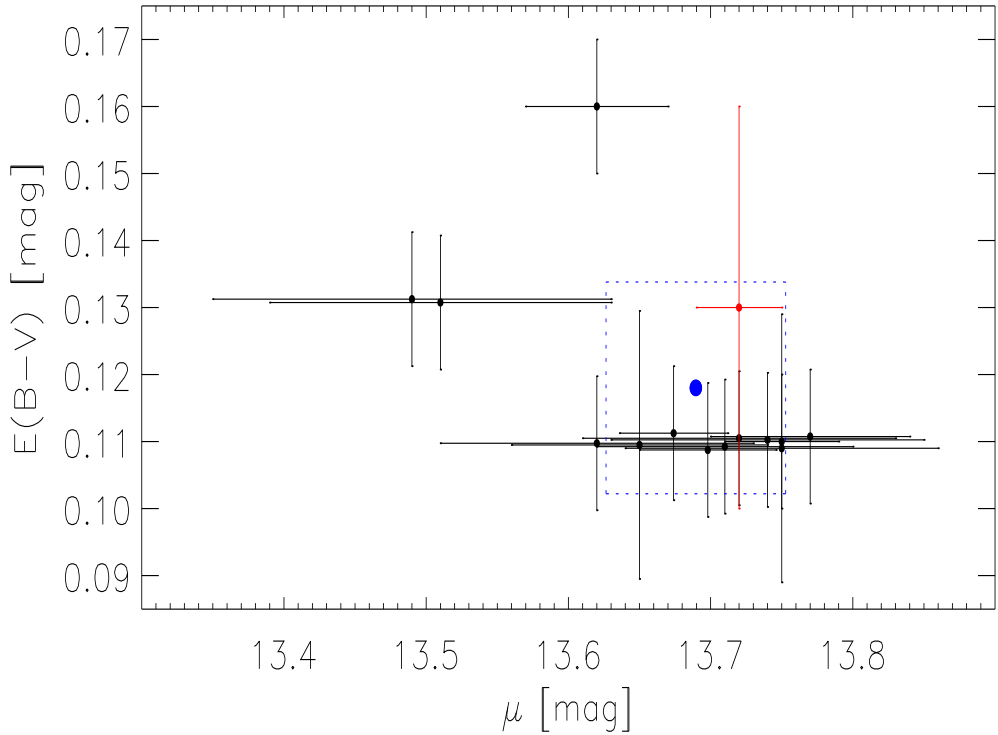}
\caption{
Comparison between cluster reddening versus cluster true distance modulus
for $\omega$ Cen based either on REDIME (red point) 
or available in the literature (black points, see also Table~3). 
The red and the black lines display either the standard deviation of the 
distributions or the the error bars. The blue dashed lines display the 
1$\sigma$ box and the mean (blue circle) of all the measurements. 
}
\label{fig:omegacen_13}
\end{figure*}

Data plotted in Fig.~\ref{fig:omegacen_13} display the comparison between recent 
estimates of both true distance modulus and reddening to $\omega$ Cen. Note that 
we decided to use this plane, since absolute distance estimates available in the 
literature are typically correlated either with the estimate of the cluster 
reddening or with the adopted cluster reddening and/or with the adopted 
reddening law.  

The mean over the literature values are (for more details see Table~3): 
$\mu$=13.690$\pm$0.018(error on the mean)$\pm$0.063 (standard deviation) mag   
and 
E($B-V$)=0.118$\pm$0.003(error on the mean)$\pm$0.016 (standard deviation) mag. 
The dashed blue box shows the 1$\sigma$ region in $\mu$ and in E($B-V$), 
while the red point and the red lines display the REDIME estimates 
and their standard deviations. Note that in several of the quoted papers 
the authors do not provide an error concerning the estimated/adopted 
cluster reddening. We adopted for these estimates a conservative error of 
0.01 mag.

There is evidence that the current true distance 
modulus is a few hundredths of a magnitudes larger than suggested in the 
literature. The same outcome also applies to the current estimate of the mean 
cluster reddening that is $\sim0.01$ mag larger than the literature values. 
However, both the cluster true distance modulus and the cluster reddening 
agree within 1$\sigma$.

\section{Summary and final remarks}\label{chapt_final}

We took advantage of the accurate and homogeneous optical ($BVI$) and NIR ($JHK$) 
mean magnitudes for RRLs in $\omega$ Cen to develop a new algorithm (REDIME) to 
fully exploit the use of RRLs as distance indicators and tracers of old 
stellar populations. The main reason why we selected $\omega$ Cen is because 
its stellar content is affected by a well known spread in metal content 
\citep{johnson_e_pilachowski2010}. Moreover, there is evidence of a mild 
variation in cluster reddening when moving across the body of the cluster. 
This means that $\omega$ Cen is a solid laboratory to evaluate the accuracy 
of the intrinsic parameters (distance, metallicity, reddening) for individual RRLs.  

We found that we cannot solve simultaneously for the three unknown parameters 
(distance, metallicity, reddening) because the adopted optical and NIR mean 
magnitudes display similar metallicity dependencies. This is the reason why 
we developed a new algorithm (REDIME) based on two steps. In the first step, 
we took advantage of the theoretical and empirical evidence that the $V$,$B-I$ 
Period-Wesenheit relation for RRLs is minimally affected by the metallicity. 
On the basis of this individual estimates of the metallicity we provided a 
preliminary estimate of the cluster distance and of the cluster reddening 
by using the same approach adopted by \citet{inno2016} for Large Magellanic 
Cloud Cepheids. In the second step, we used the NIR ($J$,$H$,$K$) PL relations 
to provide new individual metallicity estimates together with optical and 
NIR mean magnitudes to simultaneously estimate true distance modulus and 
cluster reddening. The main results of our approach are the following:\\ 

{\em i)--Metallicity distribution--} 
The metallicity distribution shows a well defined peak at 
[Fe/H]=--1.98$\pm$0.04. The spread in iron abundance is of the 
order of 0.54 dex and its is intrinsic, i.e. it is not dominated by 
uncertainties on individual measurements. There is evidence of a 
metal-intermediate group of RRLs located at [Fe/H]$\sim$--1.5 together 
with a minor tail of very metal-poor ([Fe/H]$\le$--2.3) objects.    
The comparison with metallicity distributions available in the literature 
shows that the current distribution is systematically more metal-poor 
than the spectroscopic measurements (67 objects in common) provided by 
\citet{sollima06a} and with the spectro-photometric measurements (119 
objects in common) provided by \citet{rey2000}. The differences are of the order 
of 0.3 dex (1$\sigma$). In this context it is worth mentioning that a systematic error 
of the order of --0.06,--0.07 mag in the cluster true distance modulus,
based on predicted PLZ and mean magnitude--metallicity relations, would bring 
in quite good agreement the current metallicity distribution with similar 
distributions available in the literature. The spectroscopic sample is only 
based on 67 RRLs, it is clear that larger and homogeneous samples are mandatory 
to further support the spectroscopic route to determine the accuracy and the 
precision of the true distance modulus. \\

{\em ii)--True distance modulus distribution--} 
The true distance moduli we estimated for RRc, RRab and global sample agree 
quite well with each other and the mean is $\mu$=13.720$\pm$0.002$\pm$0.030 mag.    
The quoted errors do not account for uncertainties in the zero-point. Note that the accuracy 
of the current cluster true distance modulus mainly depends on the NIR bands ($J$,$H$,$K$), 
since the slope of both PL and PW relations increases, while the standard deviation 
decreases when moving towards longer wavelengths. In passing, we also note that REDIME 
relies on multiple optical and NIR mean magnitudes. This means that REDIME simultaneously 
takes account of the optical/NIR intrinsic color variation of RRLs.  

The accuracy of the five field RRLs for which the trigonometric parallax was measured 
using the Fine Guide Sensor on board of the Hubble Space Telescope does not allow us 
to improve the current RRL distance scale. It is clear that Gaia is going to play 
a crucial role in this issue. Because the number of RRLs for which the trigonometric 
parallax is going to have an accuracy better than 1$\mu$as is two/three
orders of magnitude larger than the current ones. However, only a minor fraction of 
them have already iron abundances based on high resolution spectra 
\citep{chadid2017,magurno2018}. The bulk of metallicity estimates of field RRLs 
are still based either on medium resolution spectra or on the $\Delta$S method 
(\citealt{layden94,sesar13,kinman12,dambis14,sesar2017}; 
Fabrizio et al. 2018, in preparation).  
These metal abundances have been recently used by \citet{muraveva18b}
together with optical, near-/mid-infrared magnitudes available in the literature
and trigonometric parallaxes by Gaia DR2. They provided new Period-Luminosity-Metallicity 
relations  and found that the current parallaxes are affected by a zero-point offset 
of --0.057 mas. These finding supports previous investigations by \citet{arenou2018} 
and by \citet{sesar2017}. 

The Gaia DR2 parallaxes are systematically smaller than expected. This systematic 
error is known, it depends on several parameters (sky distribution, distance color, 
number of measurements) and it is $\sim$--0.029 mas when compared with position of more 
than 500,000 Active Galactic Nuclei \citep{lindegren2018}. In this context it is worth 
mentioning that similar analyses, but based on different stellar tracers, provide zero-point 
offsets similar to RRLs (classical Cepheids$\sim$--0.046 mas, \citealt{riess2018}
$\sim$--0.049 mas, \citealt{groenewegen2018};  red 
giants observed by KEPLER$\sim$--0.053 mas, \citealt{zinn2018}). These circumstantial evidence 
indicates that the near future scenario is very promising, but we need a few more years 
to nail down systematics in trigonometric distances.      
        
Note that Gaia is also going to provide metallicity estimates based either on spectroscopy or 
spectrophotometry for variable stars, but a significant sample would not be anticipated until 
DR4. The brightest RRL is the prototype RR Lyr itself ($m_V$=7.68 mag) and they become fainter 
than $m_V\sim$20-21 mag in the outskirts of the Galactic Halo.\\

{\em iii)--Reddening distribution--} 
We found that the cluster reddening distribution based on RRc, RRab and global 
sample agree quite well with each other and the mean is 
E($B-V$)=0.132$\pm$0.002$\pm$0.028~mag. The standard deviations of the three distributions 
are also quite similar with the distribution based on RRc being slightly narrower 
and more symmetric. The accuracy of the current reddening estimates mainly rely 
on optical bands ($BVI$) with the NIR ($JHK$) bands playing a minor role, since the slope 
of the reddening law is quite constant in this wavelength regime. The current 
cluster reddening estimates agree quite well with similar estimates recently 
provided by Gaia. We also found that the reddening changes by more than a 
factor of two on spatial scales of the order of arcminutes.  
The quoted cluster true distance modulus and cluster reddening agree within 1$\sigma$  
with similar estimates available in the literature. \\ 

{\em iv)--Metallicity dependence--} 
Preliminary empirical evidence suggest the expected correlation between optical 
magnitudes ($B$,$V$) and metal content. The current sample cover more than 1.5 dex 
and the difference in magnitude is roughly half magnitude. The same outcome applies to 
the PL relations, but the impact of the metal content at fixed period is, as expected, 
milder.  \\   

REDIME seems a very promising approach to constrain intrinsic parameters of both 
field and cluster RRLs. 
%
This working hypothesis is further strengthened by the evidence that
REDIME provides accurate estimates of metallicity, true distance modulus
and reddening for Blazkho RRLs once accurate optical/NIR mean magnitudes
are available.  In passing we note that 26 out of the 170 RRLs adopted in
this investigation are candidate Blazkho RRLs. 
The improvement of individual distances provide the opportunity 
to improve the accuracy of both metal content, reddening and possibly helium content 
\citep{marconi2018}. New and accurate spectroscopic measurements together with 
Gaia parallaxes will provide the opportunity to calibrate new optical/NIR PLZ and 
PWZ relations, and in turn, the opportunity to apply REDIME in Local Group galaxies.  

It goes without saying that REDIME was also developed to take advantage of the 
time series in six ($ugrizy$) different bands that will be collected by LSST 
for resolved stellar populations in Local Group and in Local Volume galaxies
\citep{ivezic2012}. There are solid reasons to believe that this photometric 
system is going to provide very accurate reddenings and metallicity estimates, 
but detailed simulations are required to characterize this photometric system 
for RRLs. Solid clues on the accuracy of the LSST photometric system  
in constraining metallicity, true distance modulus and reddening can also 
be derived using the multi-band, multi-epoch DECAM images 
for cluster \citep{vivas2017} and Bulge 
\citep{sahavivas2017} RRLs with the key advantage 
to cover the entire body of the cluster with a single or at most a few 
pointings \citep{calamida2017}.

\acknowledgements 
It is a real pleasure to thank the anonymous referee for her/his 
positive opinion concerning the content and the cut of this 
investigation, and in particular, for her/his constructive 
suggestions that improved the readibility of the paper.  
This investigation was partially supported by PRIN-INAF 2016
ACDC  (P.I.: P. Caraveo). 
M.M. was partially supported by NSF grant AST-1714534. 
This work has made use of data from the European Space Agency (ESA) 
mission Gaia (\url{https://www.cosmos.esa.int/gaia}), processed by 
the Gaia Data Processing and Analysis Consortium 
(DPAC, \url{https://www.cosmos.esa.int/web/gaia/dpac/consortium}). 
Funding for the DPAC has been provided by national institutions, 
in particular the institutions participating in the Gaia Multilateral 
Agreement. This research has made use of the GaiaPortal catalogues 
access tool, ASI - Space Science Data Center, Rome, Italy 
(\url{http://gaiaportal.ssdc.asi.it}).
We would also like to acknowledge the financial support of 
INAF (Istituto Nazionale di Astrofisica), Osservatorio Astronomico 
di Roma, ASI (Agenzia Spaziale Italiana) under contract to INAF: 
ASI 2014-049-R.0 dedicated to SSDC.
This publication makes use of data products from the Two Micron All Sky Survey, 
which is a joint project of the University of Massachusetts and the Infrared Processing 
and Analysis Center/California Institute of Technology, funded by the National 
Aeronautics and Space Administration and the National Science Foundation.
This research has made use of the USNO Image and Catalogue Archive
operated by the United States Naval Observatory, Flagstaff Station
(\url{http://www.nofs.navy.mil/data/fchpix/}). 
This research has made use of NASA's Astrophysics Data System.

\appendix
\centerline{FLOW CHART OF THE REDIME's ALGORITHM}  

%
\begin{figure*}[!htbp]
\centering
\includegraphics[width=14cm, height=16cm]{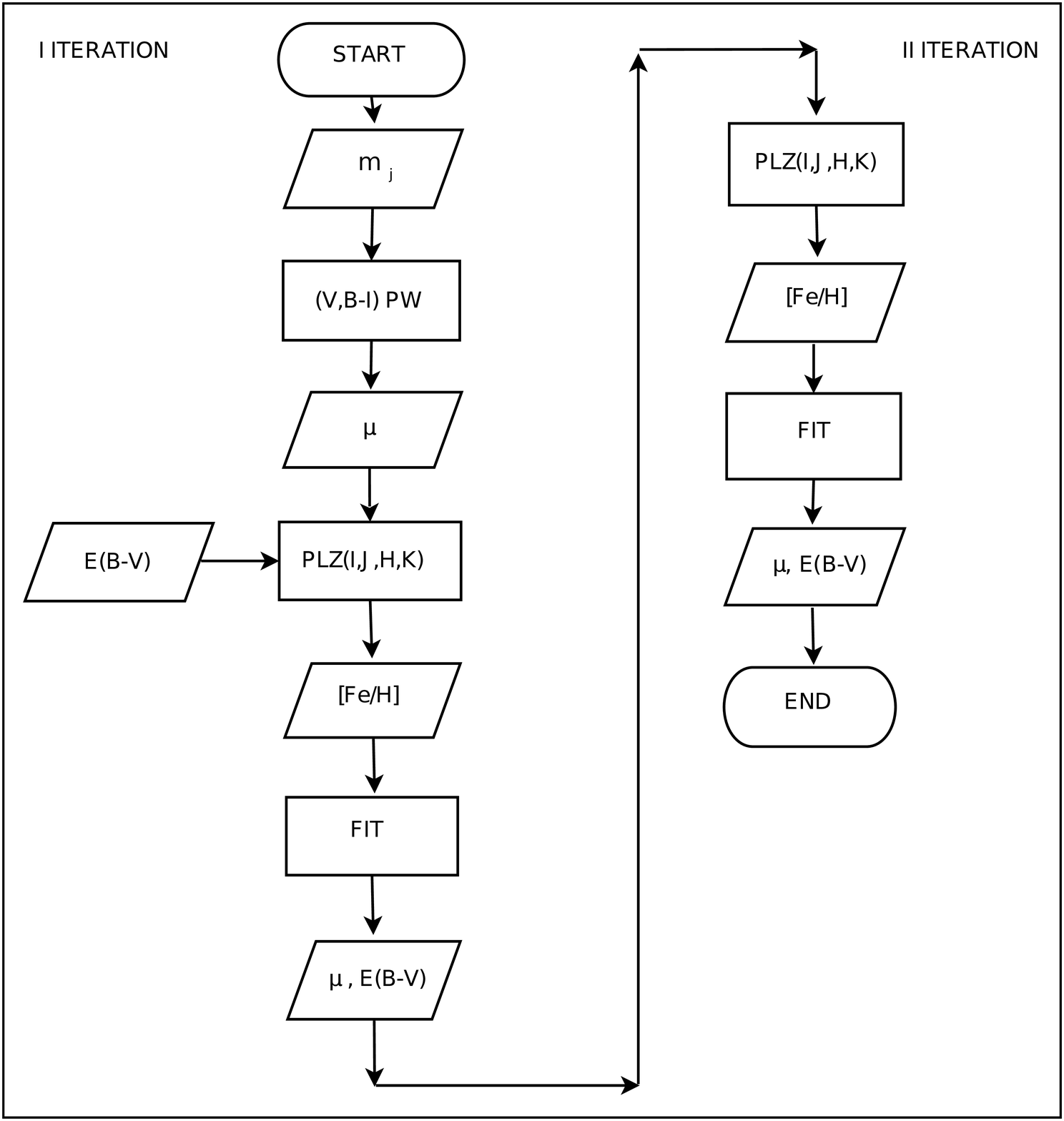}
\caption{
Left -- Flow chart of the first iteration in the REDIME's algorithm. The
algorithm is based on the six mean magnitudes available for RRLs in our
sample and the use of optical ($V$,$B-I$) PW relation to provide the first
estimate of $\omega$ Cen true distance modulus. The new distance together
with a mean cluster reddening available in the literature are used to
provide four independent estimates of  the iron abundance using the
PLZ relations in $I$,$J$,$H$,$K$  bands and equation 1. The mean metallicity
for each RRL in the sample is estimated as the mean of the estimates
based on NIR ($JHK$) PLZ relations. The new mean metallicity is used to
provide new simultaneous estimates of the true distance modulus and
reddening using four ($IJHK$) PLZ relations, two MZ ($BV$) relations
together with the nonlinear fit with the assumed reddening law
(equation 2).
Right -- The output of the first iteration, namely the true distance modulus
and the reddening are now used as an input for the second iteration. The
four $IJHK$ PLZ relations are used once again to provide new iron abundances.
The new mean value based on NIR PLZ relations is used to provide the final
simultaneous estimates of both the true distance modulus and the reddening
using the nonlinear fit dictated by equation 2. 
}
\label{fig:omegacen_1}
\end{figure*}

\bibliographystyle{aa}

\clearpage 
\setlength{\tabcolsep}{0.15cm}
\begin{deluxetable*}{l cccc  ccc ccc ccc}
 \tablewidth{0pt}
 \tabletypesize{\scriptsize}
 \tablecaption{Individual iron abundances, reddenings and true distance moduli for 
$\omega$ Cen RRLs based on REDIME.}
 \tablehead{ & & & & & & RRc\tablenotemark{e} & & & RRab\tablenotemark{e}& & & Global\tablenotemark{e} &  \\
  ID\tablenotemark{a} & Mode\tablenotemark{b} & RA\tablenotemark{c} & DEC\tablenotemark{c} & NM\tablenotemark{d}& [Fe/H]\tablenotemark{f} & E($B-V$)\tablenotemark{g} & $\mu$\tablenotemark{h} & [Fe/H]\tablenotemark{f} & E($B-V$)\tablenotemark{g} & $\mu$\tablenotemark{h} & [Fe/H]\tablenotemark{f} & E($B-V$)\tablenotemark{g} & $\mu$\tablenotemark{h} \\
         &    &              &               &     & dex                     &  mag                  & mag   & dex                     &  mag                  & mag & dex                     &  mag                  & mag
}
 \startdata
 V3    &  0 & 201.483917 & -47.431750 &  6 & \ldots          &  \ldots           &  \ldots           & -1.79 $\pm$ 0.13 &  0.10 $\pm$  0.01 & 13.74 $\pm$  0.03 &  -1.79 $\pm$ 0.11 &  0.10 $\pm$  0.01 & 13.74 $\pm$  0.02  \\
 V4    &  0 & 201.553833 & -47.405389 &  6 & \ldots          &  \ldots           &  \ldots           & -2.16 $\pm$ 0.09 &  0.13 $\pm$  0.01 & 13.71 $\pm$  0.02 &  -2.12 $\pm$ 0.05 &  0.13 $\pm$  0.01 & 13.71 $\pm$  0.02  \\
 V5    &  0 & 201.576333 & -47.386944 &  6 & \ldots          &  \ldots           &  \ldots           & -1.33 $\pm$ 0.18 &  0.16 $\pm$  0.01 & 13.70 $\pm$  0.04 &  -1.26 $\pm$ 0.14 &  0.15 $\pm$  0.01 & 13.71 $\pm$  0.03  \\
 V7    &  0 & 201.754250 & -47.233389 &  6 & \ldots          &  \ldots           &  \ldots           & -2.01 $\pm$ 0.11 &  0.14 $\pm$  0.01 & 13.71 $\pm$  0.02 &  -1.99 $\pm$ 0.07 &  0.14 $\pm$  0.01 & 13.72 $\pm$  0.02  \\
 V8    &  0 & 201.951792 & -47.472444 &  6 & \ldots          &  \ldots           &  \ldots           & -1.90 $\pm$ 0.11 &  0.16 $\pm$  0.01 & 13.70 $\pm$  0.02 &  -1.84 $\pm$ 0.05 &  0.16 $\pm$  0.01 & 13.71 $\pm$  0.01  \\
 V9    &  0 & 201.498208 & -47.440139 &  6 & \ldots          &  \ldots           &  \ldots           & -1.46 $\pm$ 0.13 &  0.18 $\pm$  0.01 & 13.70 $\pm$  0.03 &  -1.40 $\pm$ 0.08 &  0.17 $\pm$  0.01 & 13.71 $\pm$  0.02  \\
 V10   &  1 & 201.529125 & -47.410306 &  6 &-1.79 $\pm$ 0.21 &  0.11 $\pm$  0.01 & 13.70 $\pm$  0.03 &  \ldots          &  \ldots           &  \ldots           &  -2.11 $\pm$ 0.12 &  0.12 $\pm$  0.01 & 13.72 $\pm$  0.03  \\
 V11   &  0 & 201.627250 & -47.383889 &  6 & \ldots          &  \ldots           &  \ldots           & -1.98 $\pm$ 0.18 &  0.13 $\pm$  0.01 & 13.69 $\pm$  0.04 &  -2.08 $\pm$ 0.05 &  0.12 $\pm$  0.01 & 13.72 $\pm$  0.02  \\
 V12   &  1 & 201.613208 & -47.401861 &  6 &-2.16 $\pm$ 0.30 &  0.12 $\pm$  0.01 & 13.76 $\pm$  0.04 &  \ldots          &  \ldots           &  \ldots           &  -2.05 $\pm$ 0.07 &  0.13 $\pm$  0.01 & 13.71 $\pm$  0.02  \\
 V13   &  0 & 201.492375 & -47.422806 &  6 & \ldots          &  \ldots           &  \ldots           & -2.13 $\pm$ 0.12 &  0.13 $\pm$  0.01 & 13.71 $\pm$  0.03 &  -2.16 $\pm$ 0.07 &  0.13 $\pm$  0.01 & 13.73 $\pm$  0.02  \\
 V14   &  1 & 201.498583 & -47.652778 &  6 &-1.33 $\pm$ 0.41 &  0.09 $\pm$  0.02 & 13.69 $\pm$  0.06 &  \ldots          &  \ldots           &  \ldots           &  -1.91 $\pm$ 0.14 &  0.11 $\pm$  0.01 & 13.73 $\pm$  0.03  \\
 V15   &  0 & 201.612833 & -47.410694 &  6 & \ldots          &  \ldots           &  \ldots           & -1.86 $\pm$ 0.28 &  0.12 $\pm$  0.01 & 13.68 $\pm$  0.05 &  -2.10 $\pm$ 0.11 &  0.12 $\pm$  0.01 & 13.72 $\pm$  0.02  \\
 V16   &  1 & 201.907125 & -47.626444 &  6 &-2.01 $\pm$ 0.17 &  0.13 $\pm$  0.01 & 13.72 $\pm$  0.03 &  \ldots          &  \ldots           &  \ldots           &  -1.90 $\pm$ 0.14 &  0.13 $\pm$  0.01 & 13.72 $\pm$  0.03  \\
 V18   &  0 & 201.937750 & -47.415833 &  6 & \ldots          &  \ldots           &  \ldots           & -2.19 $\pm$ 0.24 &  0.13 $\pm$  0.01 & 13.76 $\pm$  0.04 &  -1.94 $\pm$ 0.07 &  0.12 $\pm$  0.01 & 13.73 $\pm$  0.02  \\
 V19   &  1 & 201.875500 & -47.468278 &  6 &-1.90 $\pm$ 0.44 &  0.18 $\pm$  0.01 & 13.76 $\pm$  0.06 &  \ldots          &  \ldots           &  \ldots           &  -1.40 $\pm$ 0.09 &  0.17 $\pm$  0.01 & 13.71 $\pm$  0.02  \\
 V20   &  0 & 201.808500 & -47.468583 &  6 & \ldots          &  \ldots           &  \ldots           & -1.73 $\pm$ 0.36 &  0.15 $\pm$  0.01 & 13.64 $\pm$  0.07 &  -2.05 $\pm$ 0.10 &  0.15 $\pm$  0.01 & 13.71 $\pm$  0.02  \\
 V21   &  1 & 201.546458 & -47.433167 &  6 &-1.46 $\pm$ 0.26 &  0.05 $\pm$  0.01 & 13.75 $\pm$  0.04 &  \ldots          &  \ldots           &  \ldots           &  -1.82 $\pm$ 0.24 &  0.06 $\pm$  0.01 & 13.76 $\pm$  0.04  \\
 V22   &  1 & 201.921042 & -47.568917 &  6 &-1.98 $\pm$ 0.29 &  0.11 $\pm$  0.01 & 13.77 $\pm$  0.04 &  \ldots          &  \ldots           &  \ldots           &  -1.94 $\pm$ 0.11 &  0.12 $\pm$  0.01 & 13.73 $\pm$  0.02  \\
 V23   &  0 & 201.693667 & -47.411028 &  6 & \ldots          &  \ldots           &  \ldots           & -2.10 $\pm$ 0.73 &  0.20 $\pm$  0.02 & 13.81 $\pm$  0.12 &  -1.31 $\pm$ 0.10 &  0.17 $\pm$  0.01 & 13.70 $\pm$  0.02  \\
 V24   &  1 & 201.909708 & -47.570833 &  6 &-2.13 $\pm$ 0.55 &  0.09 $\pm$  0.02 & 13.83 $\pm$  0.07 &  \ldots          &  \ldots           &  \ldots           &  -2.04 $\pm$ 0.11 &  0.11 $\pm$  0.01 & 13.74 $\pm$  0.02  \\
 V25   &  0 & 201.606250 & -47.473278 &  6 & \ldots          &  \ldots           &  \ldots           & -1.92 $\pm$ 0.13 &  0.12 $\pm$  0.01 & 13.70 $\pm$  0.03 &  -1.99 $\pm$ 0.02 &  0.12 $\pm$  0.01 & 13.73 $\pm$  0.01  \\
 V26   &  0 & 201.598375 & -47.449972 &  6 & \ldots          &  \ldots           &  \ldots           & -1.98 $\pm$ 0.06 &  0.14 $\pm$  0.01 & 13.71 $\pm$  0.02 &  -1.98 $\pm$ 0.04 &  0.14 $\pm$  0.01 & 13.72 $\pm$  0.01  \\
 V27   &  0 & 201.608417 & -47.471417 &  6 & \ldots          &  \ldots           &  \ldots           & -1.46 $\pm$ 0.10 &  0.18 $\pm$  0.01 & 13.66 $\pm$  0.02 &  -1.50 $\pm$ 0.09 &  0.18 $\pm$  0.01 & 13.68 $\pm$  0.02  \\
 V30   &  1 & 201.566333 & -47.499056 &  6 &-1.86 $\pm$ 0.14 &  0.09 $\pm$  0.01 & 13.76 $\pm$  0.02 &  \ldots          &  \ldots           &  \ldots           &  -2.00 $\pm$ 0.06 &  0.10 $\pm$  0.01 & 13.73 $\pm$  0.02  \\
 V32   &  0 & 201.763917 & -47.360917 &  6 & \ldots          &  \ldots           &  \ldots           & -2.08 $\pm$ 0.22 &  0.12 $\pm$  0.01 & 13.76 $\pm$  0.04 &  -1.83 $\pm$ 0.07 &  0.11 $\pm$  0.01 & 13.74 $\pm$  0.02  \\
 V33   &  0 & 201.464833 & -47.485056 &  6 & \ldots          &  \ldots           &  \ldots           & -1.62 $\pm$ 0.53 &  0.13 $\pm$  0.01 & 13.63 $\pm$  0.09 &  -2.08 $\pm$ 0.11 &  0.14 $\pm$  0.01 & 13.73 $\pm$  0.02  \\
 V34   &  0 & 201.529917 & -47.553056 &  6 & \ldots          &  \ldots           &  \ldots           & -1.70 $\pm$ 0.41 &  0.13 $\pm$  0.01 & 13.65 $\pm$  0.07 &  -2.08 $\pm$ 0.05 &  0.13 $\pm$  0.01 & 13.72 $\pm$  0.01  \\
 V35   &  1 & 201.721833 & -47.376389 &  6 &-2.19 $\pm$ 0.30 &  0.12 $\pm$  0.01 & 13.77 $\pm$  0.04 &  \ldots          &  \ldots           &  \ldots           &  -2.09 $\pm$ 0.13 &  0.13 $\pm$  0.01 & 13.72 $\pm$  0.03  \\
 V36   &  1 & 201.792458 & -47.258222 &  6 &-1.73 $\pm$ 0.27 &  0.10 $\pm$  0.01 & 13.70 $\pm$  0.04 &  \ldots          &  \ldots           &  \ldots           &  -2.13 $\pm$ 0.13 &  0.12 $\pm$  0.01 & 13.73 $\pm$  0.03  \\
 V38   &  0 & 201.763500 & -47.608472 &  6 & \ldots          &  \ldots           &  \ldots           & -1.38 $\pm$ 0.66 &  0.12 $\pm$  0.01 & 13.61 $\pm$  0.10 &  -2.02 $\pm$ 0.05 &  0.13 $\pm$  0.01 & 13.72 $\pm$  0.01  \\
 V39   &  1 & 201.999250 & -47.578444 &  6 &-2.10 $\pm$ 0.58 &  0.10 $\pm$  0.02 & 13.82 $\pm$  0.08 &  \ldots          &  \ldots           &  \ldots           &  -1.79 $\pm$ 0.10 &  0.10 $\pm$  0.01 & 13.74 $\pm$  0.02  \\
 V40   &  0 & 201.602292 & -47.513000 &  6 & \ldots          &  \ldots           &  \ldots           & -1.59 $\pm$ 0.39 &  0.13 $\pm$  0.01 & 13.65 $\pm$  0.07 &  -1.94 $\pm$ 0.02 &  0.13 $\pm$  0.01 & 13.72 $\pm$  0.02  \\
 V41   &  0 & 201.755750 & -47.517278 &  6 & \ldots          &  \ldots           &  \ldots           & -2.03 $\pm$ 0.06 &  0.14 $\pm$  0.01 & 13.71 $\pm$  0.02 &  -1.96 $\pm$ 0.01 &  0.13 $\pm$  0.01 & 13.71 $\pm$  0.01  \\
 V44   &  0 & 201.593250 & -47.576611 &  6 & \ldots          &  \ldots           &  \ldots           & -1.99 $\pm$ 0.57 &  0.18 $\pm$  0.01 & 13.79 $\pm$  0.10 &  -1.37 $\pm$ 0.06 &  0.16 $\pm$  0.01 & 13.70 $\pm$  0.02  \\
 V45   &  0 & 201.378542 & -47.455833 &  6 & \ldots          &  \ldots           &  \ldots           & -1.54 $\pm$ 0.58 &  0.13 $\pm$  0.01 & 13.62 $\pm$  0.10 &  -2.06 $\pm$ 0.11 &  0.14 $\pm$  0.01 & 13.72 $\pm$  0.02  \\
 V46   &  0 & 201.375958 & -47.431028 &  6 & \ldots          &  \ldots           &  \ldots           & -1.75 $\pm$ 0.29 &  0.14 $\pm$  0.01 & 13.69 $\pm$  0.05 &  -2.00 $\pm$ 0.06 &  0.14 $\pm$  0.01 & 13.74 $\pm$  0.02  \\
 V47   &  1 & 201.485292 & -47.403444 &  6 &-1.92 $\pm$ 0.32 &  0.05 $\pm$  0.01 & 13.82 $\pm$  0.04 &  \ldots          &  \ldots           &  \ldots           &  -2.13 $\pm$ 0.17 &  0.08 $\pm$  0.01 & 13.76 $\pm$  0.03  \\
 V49   &  0 & 201.532167 & -47.632222 &  6 & \ldots          &  \ldots           &  \ldots           & -1.87 $\pm$ 0.05 &  0.17 $\pm$  0.01 & 13.70 $\pm$  0.02 &  -1.80 $\pm$ 0.02 &  0.16 $\pm$  0.01 & 13.70 $\pm$  0.01  \\
 V50   &  1 & 201.474667 & -47.460056 &  6 &-1.98 $\pm$ 0.45 &  0.14 $\pm$  0.01 & 13.77 $\pm$  0.06 &  \ldots          &  \ldots           &  \ldots           &  -1.77 $\pm$ 0.10 &  0.14 $\pm$  0.01 & 13.70 $\pm$  0.02  \\
 V51   &  0 & 201.677417 & -47.406028 &  6 & \ldots          &  \ldots           &  \ldots           & -2.12 $\pm$ 0.07 &  0.14 $\pm$  0.01 & 13.71 $\pm$  0.02 &  -2.12 $\pm$ 0.05 &  0.14 $\pm$  0.01 & 13.72 $\pm$  0.02  \\
 V52   &  0 & 201.646458 & -47.467917 &  6 & \ldots          &  \ldots           &  \ldots           & -2.10 $\pm$ 0.26 &  0.07 $\pm$  0.01 & 13.71 $\pm$  0.05 &  -2.28 $\pm$ 0.13 &  0.07 $\pm$  0.01 & 13.76 $\pm$  0.03  \\
 V54   &  0 & 201.597917 & -47.313389 &  6 & \ldots          &  \ldots           &  \ldots           & -1.91 $\pm$ 0.15 &  0.12 $\pm$  0.01 & 13.69 $\pm$  0.03 &  -2.04 $\pm$ 0.02 &  0.12 $\pm$  0.01 & 13.72 $\pm$  0.01  \\
 V56   &  0 & 201.481000 & -47.629028 &  6 & \ldots          &  \ldots           &  \ldots           & -1.99 $\pm$ 0.74 &  0.18 $\pm$  0.02 & 13.83 $\pm$  0.13 &  -1.20 $\pm$ 0.10 &  0.16 $\pm$  0.01 & 13.71 $\pm$  0.02  \\
 V57   &  0 & 201.955917 & -47.614083 &  6 & \ldots          &  \ldots           &  \ldots           & -2.04 $\pm$ 0.08 &  0.14 $\pm$  0.01 & 13.70 $\pm$  0.02 &  -2.01 $\pm$ 0.07 &  0.14 $\pm$  0.01 & 13.70 $\pm$  0.02  \\
 V58   &  1 & 201.554292 & -47.400972 &  6 &-1.46 $\pm$ 0.15 &  0.06 $\pm$  0.01 & 13.75 $\pm$  0.03 &  \ldots          &  \ldots           &  \ldots           &  -1.75 $\pm$ 0.18 &  0.08 $\pm$  0.01 & 13.75 $\pm$  0.03  \\
 V59   &  0 & 201.576708 & -47.496500 &  6 & \ldots          &  \ldots           &  \ldots           & -1.53 $\pm$ 0.17 &  0.16 $\pm$  0.01 & 13.68 $\pm$  0.04 &  -1.52 $\pm$ 0.12 &  0.15 $\pm$  0.01 & 13.70 $\pm$  0.03  \\
 V62   &  0 & 201.610708 & -47.465556 &  6 & \ldots          &  \ldots           &  \ldots           & -1.97 $\pm$ 0.17 &  0.11 $\pm$  0.01 & 13.69 $\pm$  0.03 &  -2.10 $\pm$ 0.04 &  0.11 $\pm$  0.01 & 13.73 $\pm$  0.02  \\
 V63   &  0 & 201.282792 & -47.614972 &  6 & \ldots          &  \ldots           &  \ldots           & -1.99 $\pm$ 0.09 &  0.13 $\pm$  0.01 & 13.73 $\pm$  0.02 &  -1.93 $\pm$ 0.05 &  0.13 $\pm$  0.01 & 13.73 $\pm$  0.02  \\
 V64   &  1 & 201.509000 & -47.605472 &  6 &-2.08 $\pm$ 0.14 &  0.13 $\pm$  0.01 & 13.73 $\pm$  0.02 &  \ldots          &  \ldots           &  \ldots           &  -2.02 $\pm$ 0.13 &  0.13 $\pm$  0.01 & 13.72 $\pm$  0.03  \\
 V66   &  1 & 201.637583 & -47.373806 &  6 &-1.62 $\pm$ 0.26 &  0.10 $\pm$  0.01 & 13.70 $\pm$  0.04 &  \ldots          &  \ldots           &  \ldots           &  -2.11 $\pm$ 0.11 &  0.13 $\pm$  0.01 & 13.72 $\pm$  0.03  \\
 V67   &  0 & 201.619000 & -47.313139 &  6 & \ldots          &  \ldots           &  \ldots           & -1.42 $\pm$ 0.10 &  0.15 $\pm$  0.01 & 13.71 $\pm$  0.02 &  -1.43 $\pm$ 0.06 &  0.14 $\pm$  0.01 & 13.72 $\pm$  0.02  \\
 V69   &  0 & 201.295542 & -47.625917 &  6 & \ldots          &  \ldots           &  \ldots           & -2.11 $\pm$ 0.27 &  0.13 $\pm$  0.03 & 13.76 $\pm$  0.05 &  -1.82 $\pm$ 0.07 &  0.12 $\pm$  0.03 & 13.72 $\pm$  0.03  \\
 V70   &  1 & 201.865625 & -47.562028 &  6 &-1.70 $\pm$ 0.24 &  0.10 $\pm$  0.01 & 13.71 $\pm$  0.04 &  \ldots          &  \ldots           &  \ldots           &  -2.12 $\pm$ 0.10 &  0.11 $\pm$  0.01 & 13.73 $\pm$  0.02  \\
 V71   &  1 & 201.783583 & -47.464528 &  6 &-1.38 $\pm$ 0.42 &  0.10 $\pm$  0.01 & 13.66 $\pm$  0.06 &  \ldots          &  \ldots           &  \ldots           &  -1.90 $\pm$ 0.09 &  0.12 $\pm$  0.01 & 13.72 $\pm$  0.02  \\
 V72   &  1 & 201.887583 & -47.273000 &  6 &-1.59 $\pm$ 0.27 &  0.09 $\pm$  0.01 & 13.71 $\pm$  0.04 &  \ldots          &  \ldots           &  \ldots           &  -2.00 $\pm$ 0.15 &  0.11 $\pm$  0.01 & 13.74 $\pm$  0.03  \\
 V73   &  0 & 201.473542 & -47.269639 &  6 & \ldots          &  \ldots           &  \ldots           & -2.02 $\pm$ 0.15 &  0.21 $\pm$  0.04 & 13.70 $\pm$  0.04 &  -1.87 $\pm$ 0.06 &  0.20 $\pm$  0.04 & 13.69 $\pm$  0.03  \\
 V74   &  0 & 201.780208 & -47.292889 &  6 & \ldots          &  \ldots           &  \ldots           & -1.62 $\pm$ 0.28 &  0.14 $\pm$  0.01 & 13.66 $\pm$  0.05 &  -1.81 $\pm$ 0.08 &  0.13 $\pm$  0.01 & 13.71 $\pm$  0.02  \\
 V75   &  1 & 201.832042 & -47.313083 &  6 &-2.03 $\pm$ 0.11 &  0.12 $\pm$  0.01 & 13.72 $\pm$  0.02 &  \ldots          &  \ldots           &  \ldots           &  -2.31 $\pm$ 0.12 &  0.14 $\pm$  0.01 & 13.71 $\pm$  0.03  \\
 V76   &  1 & 201.738667 & -47.335556 &  6 &-1.99 $\pm$ 0.26 &  0.11 $\pm$  0.01 & 13.73 $\pm$  0.04 &  \ldots          &  \ldots           &  \ldots           &  -1.98 $\pm$ 0.26 &  0.11 $\pm$  0.01 & 13.74 $\pm$  0.04  \\
 V77   &  1 & 201.836958 & -47.368389 &  6 &-1.54 $\pm$ 0.28 &  0.10 $\pm$  0.01 & 13.69 $\pm$  0.04 &  \ldots          &  \ldots           &  \ldots           &  -2.12 $\pm$ 0.10 &  0.13 $\pm$  0.01 & 13.71 $\pm$  0.03  \\
 V79   &  0 & 202.104458 & -47.490250 &  6 & \ldots          &  \ldots           &  \ldots           & -1.85 $\pm$ 0.08 &  0.14 $\pm$  0.01 & 13.69 $\pm$  0.02 &  -1.80 $\pm$ 0.04 &  0.14 $\pm$  0.01 & 13.70 $\pm$  0.01  \\
 V81   &  1 & 201.902667 & -47.413583 &  6 &-1.75 $\pm$ 0.10 &  0.11 $\pm$  0.01 & 13.72 $\pm$  0.02 &  \ldots          &  \ldots           &  \ldots           &  -2.02 $\pm$ 0.10 &  0.13 $\pm$  0.01 & 13.73 $\pm$  0.02  \\
 V82   &  1 & 201.898250 & -47.441917 &  6 &-1.87 $\pm$ 0.56 &  0.13 $\pm$  0.02 & 13.64 $\pm$  0.08 &  \ldots          &  \ldots           &  \ldots           &  -2.34 $\pm$ 0.16 &  0.14 $\pm$  0.01 & 13.72 $\pm$  0.03  \\
 V83   &  1 & 201.785125 & -47.359611 &  6 &-2.12 $\pm$ 0.19 &  0.12 $\pm$  0.01 & 13.74 $\pm$  0.03 &  \ldots          &  \ldots           &  \ldots           &  -2.03 $\pm$ 0.13 &  0.12 $\pm$  0.01 & 13.72 $\pm$  0.03  \\
 V85   &  0 & 201.277458 & -47.392694 &  6 & \ldots          &  \ldots           &  \ldots           & -1.54 $\pm$ 0.29 &  0.12 $\pm$  0.01 & 13.67 $\pm$  0.05 &  -1.81 $\pm$ 0.04 &  0.12 $\pm$  0.01 & 13.72 $\pm$  0.01  \\
 V86   &  0 & 201.813208 & -47.436583 &  6 & \ldots          &  \ldots           &  \ldots           & -2.16 $\pm$ 0.08 &  0.15 $\pm$  0.01 & 13.70 $\pm$  0.02 &  -2.06 $\pm$ 0.05 &  0.15 $\pm$  0.01 & 13.70 $\pm$  0.02  \\
 V87   &  1 & 201.739458 & -47.426611 &  6 &-2.10 $\pm$ 0.26 &  0.13 $\pm$  0.01 & 13.75 $\pm$  0.04 &  \ldots          &  \ldots           &  \ldots           &  -2.06 $\pm$ 0.05 &  0.14 $\pm$  0.01 & 13.71 $\pm$  0.02  \\
 V88   &  0 & 201.732917 & -47.421306 &  6 & \ldots          &  \ldots           &  \ldots           & -2.31 $\pm$ 0.62 &  0.13 $\pm$  0.02 & 13.83 $\pm$  0.11 &  -1.68 $\pm$ 0.08 &  0.11 $\pm$  0.01 & 13.73 $\pm$  0.02  \\
 V89   &  1 & 201.691500 & -47.433694 &  6 &-1.91 $\pm$ 0.15 &  0.16 $\pm$  0.01 & 13.69 $\pm$  0.03 &  \ldots          &  \ldots           &  \ldots           &  -2.14 $\pm$ 0.03 &  0.17 $\pm$  0.01 & 13.70 $\pm$  0.02  \\
 V90   &  0 & 201.690500 & -47.439917 &  6 & \ldots          &  \ldots           &  \ldots           & -2.06 $\pm$ 0.16 &  0.13 $\pm$  0.01 & 13.73 $\pm$  0.03 &  -2.04 $\pm$ 0.13 &  0.12 $\pm$  0.01 & 13.75 $\pm$  0.03  \\
 V91   &  0 & 201.710750 & -47.437722 &  6 & \ldots          &  \ldots           &  \ldots           & -1.25 $\pm$ 0.92 &  0.10 $\pm$  0.02 & 13.57 $\pm$  0.14 &  -2.17 $\pm$ 0.03 &  0.12 $\pm$  0.01 & 13.72 $\pm$  0.02  \\
 V94   &  1 & 201.487792 & -47.379583 &  6 &-1.99 $\pm$ 0.33 &  0.17 $\pm$  0.01 & 13.72 $\pm$  0.05 &  \ldots          &  \ldots           &  \ldots           &  -1.40 $\pm$ 0.16 &  0.14 $\pm$  0.01 & 13.71 $\pm$  0.03  \\
 V95   &  1 & 201.353708 & -47.481389 &  6 &-2.04 $\pm$ 0.29 &  0.13 $\pm$  0.01 & 13.75 $\pm$  0.04 &  \ldots          &  \ldots           &  \ldots           &  -2.03 $\pm$ 0.14 &  0.14 $\pm$  0.01 & 13.71 $\pm$  0.03  \\
 V96   &  0 & 201.663625 & -47.450944 &  6 & \ldots          &  \ldots           &  \ldots           & -2.02 $\pm$ 0.32 &  0.14 $\pm$  0.01 & 13.65 $\pm$  0.06 &  -2.30 $\pm$ 0.06 &  0.14 $\pm$  0.01 & 13.71 $\pm$  0.02  \\
 V97   &  0 & 201.785375 & -47.425417 &  6 & \ldots          &  \ldots           &  \ldots           & -1.53 $\pm$ 0.53 &  0.12 $\pm$  0.01 & 13.63 $\pm$  0.09 &  -2.04 $\pm$ 0.03 &  0.12 $\pm$  0.01 & 13.73 $\pm$  0.01  \\
 V98   &  1 & 201.774333 & -47.449222 &  6 &-1.53 $\pm$ 0.12 &  0.17 $\pm$  0.01 & 13.68 $\pm$  0.02 &  \ldots          &  \ldots           &  \ldots           &  -1.39 $\pm$ 0.12 &  0.16 $\pm$  0.01 & 13.70 $\pm$  0.03  \\
 V99   &  0 & 201.758917 & -47.463722 &  6 & \ldots          &  \ldots           &  \ldots           & -1.58 $\pm$ 0.74 &  0.07 $\pm$  0.02 & 13.63 $\pm$  0.12 &  -2.30 $\pm$ 0.09 &  0.08 $\pm$  0.01 & 13.75 $\pm$  0.02  \\
 V100  &  0 & 201.766750 & -47.459417 &  6 & \ldots          &  \ldots           &  \ldots           & -2.13 $\pm$ 0.69 &  0.15 $\pm$  0.02 & 13.80 $\pm$  0.12 &  -1.39 $\pm$ 0.09 &  0.14 $\pm$  0.01 & 13.70 $\pm$  0.02  \\
 V101  &  1 & 201.875917 & -47.497694 &  6 &-1.97 $\pm$ 0.14 &  0.13 $\pm$  0.01 & 13.73 $\pm$  0.03 &  \ldots          &  \ldots           &  \ldots           &  -1.94 $\pm$ 0.14 &  0.13 $\pm$  0.01 & 13.72 $\pm$  0.03  \\
 V102  &  0 & 201.842042 & -47.503611 &  6 & \ldots          &  \ldots           &  \ldots           & -1.93 $\pm$ 0.11 &  0.15 $\pm$  0.01 & 13.69 $\pm$  0.03 &  -2.02 $\pm$ 0.01 &  0.15 $\pm$  0.01 & 13.71 $\pm$  0.02  \\
 V103  &  1 & 201.809458 & -47.476917 &  6 &-1.99 $\pm$ 0.26 &  0.12 $\pm$  0.01 & 13.68 $\pm$  0.04 &  \ldots          &  \ldots           &  \ldots           &  -2.16 $\pm$ 0.12 &  0.12 $\pm$  0.01 & 13.72 $\pm$  0.03  \\
 V104  &  0 & 202.032500 & -47.562472 &  6 & \ldots          &  \ldots           &  \ldots           & -2.00 $\pm$ 0.40 &  0.14 $\pm$  0.01 & 13.78 $\pm$  0.07 &  -1.60 $\pm$ 0.04 &  0.13 $\pm$  0.01 & 13.72 $\pm$  0.01  \\
 V105  &  1 & 201.941792 & -47.545639 &  6 &-1.42 $\pm$ 0.11 &  0.14 $\pm$  0.01 & 13.71 $\pm$  0.02 &  \ldots          &  \ldots           &  \ldots           &  -1.48 $\pm$ 0.11 &  0.15 $\pm$  0.01 & 13.71 $\pm$  0.02  \\
 V106  &  0 & 201.746500 & -47.470278 &  6 & \ldots          &  \ldots           &  \ldots           & -1.86 $\pm$ 0.26 &  0.13 $\pm$  0.01 & 13.68 $\pm$  0.05 &  -2.05 $\pm$ 0.07 &  0.13 $\pm$  0.01 & 13.73 $\pm$  0.02  \\
 V107  &  0 & 201.808458 & -47.516250 &  6 & \ldots          &  \ldots           &  \ldots           & -1.48 $\pm$ 0.10 &  0.17 $\pm$  0.01 & 13.68 $\pm$  0.02 &  -1.51 $\pm$ 0.02 &  0.17 $\pm$  0.01 & 13.70 $\pm$  0.01  \\
 V108  &  0 & 201.769458 & -47.490611 &  6 & \ldots          &  \ldots           &  \ldots           & -1.85 $\pm$ 0.31 &  0.14 $\pm$  0.01 & 13.66 $\pm$  0.06 &  -2.12 $\pm$ 0.01 &  0.14 $\pm$  0.01 & 13.72 $\pm$  0.01  \\
 V109  &  0 & 201.756375 & -47.493639 &  6 & \ldots          &  \ldots           &  \ldots           & -1.93 $\pm$ 0.13 &  0.11 $\pm$  0.01 & 13.71 $\pm$  0.03 &  -2.03 $\pm$ 0.02 &  0.11 $\pm$  0.01 & 13.74 $\pm$  0.01  \\
 V110  &  1 & 201.758542 & -47.502000 &  6 &-2.11 $\pm$ 0.34 &  0.13 $\pm$  0.01 & 13.76 $\pm$  0.05 &  \ldots          &  \ldots           &  \ldots           &  -1.79 $\pm$ 0.13 &  0.13 $\pm$  0.01 & 13.73 $\pm$  0.03  \\
 V111  &  0 & 201.704125 & -47.477944 &  6 & \ldots          &  \ldots           &  \ldots           & -1.80 $\pm$ 0.30 &  0.15 $\pm$  0.01 & 13.65 $\pm$  0.06 &  -2.03 $\pm$ 0.18 &  0.15 $\pm$  0.01 & 13.70 $\pm$  0.04  \\
 V112  &  0 & 201.726000 & -47.506583 &  6 & \ldots          &  \ldots           &  \ldots           & -1.82 $\pm$ 0.35 &  0.08 $\pm$  0.01 & 13.79 $\pm$  0.06 &  -1.42 $\pm$ 0.17 &  0.07 $\pm$  0.01 & 13.75 $\pm$  0.04  \\
 V113  &  0 & 201.734583 & -47.530000 &  6 & \ldots          &  \ldots           &  \ldots           & -1.92 $\pm$ 0.15 &  0.15 $\pm$  0.01 & 13.73 $\pm$  0.03 &  -1.75 $\pm$ 0.06 &  0.14 $\pm$  0.01 & 13.72 $\pm$  0.02  \\
 V114  &  0 & 201.708792 & -47.505972 &  6 & \ldots          &  \ldots           &  \ldots           & -1.87 $\pm$ 0.16 &  0.13 $\pm$  0.01 & 13.71 $\pm$  0.03 &  -1.85 $\pm$ 0.14 &  0.12 $\pm$  0.01 & 13.72 $\pm$  0.03  \\
 V115  &  0 & 201.551125 & -47.571694 &  6 & \ldots          &  \ldots           &  \ldots           & -2.04 $\pm$ 0.09 &  0.14 $\pm$  0.01 & 13.71 $\pm$  0.02 &  -2.00 $\pm$ 0.06 &  0.14 $\pm$  0.01 & 13.71 $\pm$  0.02  \\
 V116  &  0 & 201.647833 & -47.468694 &  6 & \ldots          &  \ldots           &  \ldots           & -1.98 $\pm$ 0.28 &  0.12 $\pm$  0.01 & 13.74 $\pm$  0.05 &  -1.74 $\pm$ 0.15 &  0.11 $\pm$  0.01 & 13.71 $\pm$  0.03  \\
 V117  &  1 & 201.582875 & -47.489361 &  6 &-2.02 $\pm$ 0.34 &  0.10 $\pm$  0.01 & 13.79 $\pm$  0.05 &  \ldots          &  \ldots           &  \ldots           &  -2.01 $\pm$ 0.12 &  0.12 $\pm$  0.01 & 13.73 $\pm$  0.02  \\
 V118  &  0 & 201.668917 & -47.505417 &  6 & \ldots          &  \ldots           &  \ldots           & -1.81 $\pm$ 1.42 &  0.20 $\pm$  0.03 & 13.41 $\pm$  0.20 &  -3.19 $\pm$ 0.08 &  0.22 $\pm$  0.01 & 13.66 $\pm$  0.02  \\
 V119  &  1 & 201.659500 & -47.521778 &  6 &-1.62 $\pm$ 0.49 &  0.14 $\pm$  0.01 & 13.64 $\pm$  0.07 &  \ldots          &  \ldots           &  \ldots           &  -1.96 $\pm$ 0.10 &  0.14 $\pm$  0.01 & 13.71 $\pm$  0.02  \\
 V120  &  0 & 201.606333 & -47.547000 &  6 & \ldots          &  \ldots           &  \ldots           & -1.84 $\pm$ 0.32 &  0.18 $\pm$  0.01 & 13.73 $\pm$  0.06 &  -1.48 $\pm$ 0.09 &  0.17 $\pm$  0.01 & 13.69 $\pm$  0.02  \\
 V121  &  1 & 201.617333 & -47.530861 &  6 &-1.85 $\pm$ 0.66 &  0.14 $\pm$  0.02 & 13.61 $\pm$  0.09 &  \ldots          &  \ldots           &  \ldots           &  -2.28 $\pm$ 0.12 &  0.14 $\pm$  0.01 & 13.71 $\pm$  0.03  \\
 V122  &  0 & 201.626250 & -47.550750 &  6 & \ldots          &  \ldots           &  \ldots           & -1.82 $\pm$ 0.31 &  0.15 $\pm$  0.01 & 13.65 $\pm$  0.06 &  -2.08 $\pm$ 0.08 &  0.15 $\pm$  0.01 & 13.71 $\pm$  0.02  \\
 V123  &  1 & 201.712792 & -47.620389 &  6 &-1.54 $\pm$ 0.31 &  0.06 $\pm$  0.01 & 13.82 $\pm$  0.05 &  \ldots          &  \ldots           &  \ldots           &  -1.78 $\pm$ 0.14 &  0.09 $\pm$  0.01 & 13.75 $\pm$  0.03  \\
 V124  &  1 & 201.726583 & -47.652139 &  6 &-2.16 $\pm$ 0.22 &  0.14 $\pm$  0.01 & 13.75 $\pm$  0.03 &  \ldots          &  \ldots           &  \ldots           &  -1.95 $\pm$ 0.14 &  0.13 $\pm$  0.01 & 13.74 $\pm$  0.03  \\
 V125  &  0 & 201.704042 & -47.684361 &  6 & \ldots          &  \ldots           &  \ldots           & -2.41 $\pm$ 0.51 &  0.17 $\pm$  0.01 & 13.79 $\pm$  0.10 &  -1.87 $\pm$ 0.08 &  0.16 $\pm$  0.01 & 13.71 $\pm$  0.04  \\
 V126  &  1 & 202.033875 & -47.679583 &  6 &-2.31 $\pm$ 0.53 &  0.12 $\pm$  0.02 & 13.79 $\pm$  0.08 &  \ldots          &  \ldots           &  \ldots           &  -1.83 $\pm$ 0.09 &  0.12 $\pm$  0.01 & 13.73 $\pm$  0.02  \\
 V127  &  1 & 201.330917 & -47.477111 &  6 &-2.06 $\pm$ 0.15 &  0.14 $\pm$  0.01 & 13.71 $\pm$  0.03 &  \ldots          &  \ldots           &  \ldots           &  -1.89 $\pm$ 0.16 &  0.13 $\pm$  0.01 & 13.72 $\pm$  0.03  \\
 V128  &  0 & 201.573875 & -47.503806 &  6 & \ldots          &  \ldots           &  \ldots           & -1.96 $\pm$ 0.27 &  0.11 $\pm$  0.02 & 13.69 $\pm$  0.06 &  -2.22 $\pm$ 0.07 &  0.11 $\pm$  0.02 & 13.73 $\pm$  0.04  \\
 V130  &  0 & 201.541667 & -47.227778 &  6 & \ldots          &  \ldots           &  \ldots           & -1.83 $\pm$ 0.30 &  0.18 $\pm$  0.01 & 13.74 $\pm$  0.05 &  -1.46 $\pm$ 0.06 &  0.17 $\pm$  0.01 & 13.70 $\pm$  0.02  \\
 V131  &  1 & 201.625208 & -47.494806 &  6 &-1.25 $\pm$ 0.80 &  0.09 $\pm$  0.02 & 13.62 $\pm$  0.09 &  \ldots          &  \ldots           &  \ldots           &  -2.22 $\pm$ 0.07 &  0.12 $\pm$  0.01 & 13.72 $\pm$  0.02  \\
 V132  &  0 & 201.663292 & -47.486167 &  6 & \ldots          &  \ldots           &  \ldots           & -2.10 $\pm$ 0.16 &  0.12 $\pm$  0.01 & 13.68 $\pm$  0.04 &  -2.21 $\pm$ 0.08 &  0.12 $\pm$  0.01 & 13.72 $\pm$  0.02  \\
 V134  &  0 & 201.305458 & -47.207889 &  6 & \ldots          &  \ldots           &  \ldots           & -1.84 $\pm$ 0.19 &  0.15 $\pm$  0.01 & 13.66 $\pm$  0.04 &  -1.89 $\pm$ 0.13 &  0.14 $\pm$  0.01 & 13.69 $\pm$  0.03  \\
 V135  &  0 & 201.616958 & -47.488444 &  6 & \ldots          &  \ldots           &  \ldots           & -1.71 $\pm$ 0.92 &  0.13 $\pm$  0.02 & 13.55 $\pm$  0.14 &  -2.54 $\pm$ 0.28 &  0.14 $\pm$  0.01 & 13.70 $\pm$  0.06  \\
 V136  &  1 & 201.629458 & -47.461389 &  6 &-2.02 $\pm$ 0.10 &  0.07 $\pm$  0.01 & 13.74 $\pm$  0.02 &  \ldots          &  \ldots           &  \ldots           &  -2.23 $\pm$ 0.12 &  0.09 $\pm$  0.01 & 13.74 $\pm$  0.03  \\
 V137  &  1 & 201.631333 & -47.451333 &  6 &-1.53 $\pm$ 0.69 &  0.13 $\pm$  0.02 & 13.62 $\pm$  0.09 &  \ldots          &  \ldots           &  \ldots           &  -2.17 $\pm$ 0.07 &  0.14 $\pm$  0.01 & 13.71 $\pm$  0.02  \\
 V139  &  0 & 201.657208 & -47.459889 &  6 & \ldots          &  \ldots           &  \ldots           & -2.02 $\pm$ 1.37 &  0.18 $\pm$  0.03 & 13.43 $\pm$  0.20 &  -3.37 $\pm$ 0.07 &  0.20 $\pm$  0.01 & 13.67 $\pm$  0.02  \\
 V140  &  0 & 201.675458 & -47.502000 &  6 & \ldots          &  \ldots           &  \ldots           & -2.08 $\pm$ 0.11 &  0.08 $\pm$  0.01 & 13.75 $\pm$  0.03 &  -2.02 $\pm$ 0.13 &  0.07 $\pm$  0.01 & 13.76 $\pm$  0.03  \\
 V141  &  0 & 201.670333 & -47.491194 &  6 & \ldots          &  \ldots           &  \ldots           & -2.16 $\pm$ 0.16 &  0.15 $\pm$  0.01 & 13.68 $\pm$  0.03 &  -2.30 $\pm$ 0.03 &  0.15 $\pm$  0.01 & 13.71 $\pm$  0.02  \\
 V143  &  0 & 201.677458 & -47.458083 &  6 & \ldots          &  \ldots           &  \ldots           & -1.71 $\pm$ 0.63 &  0.12 $\pm$  0.03 & 13.60 $\pm$  0.10 &  -2.32 $\pm$ 0.15 &  0.12 $\pm$  0.02 & 13.71 $\pm$  0.03  \\
 V144  &  0 & 201.679292 & -47.471722 &  6 & \ldots          &  \ldots           &  \ldots           & -1.77 $\pm$ 0.38 &  0.14 $\pm$  0.01 & 13.65 $\pm$  0.07 &  -2.13 $\pm$ 0.13 &  0.14 $\pm$  0.01 & 13.71 $\pm$  0.03  \\
 V145  &  1 & 201.713417 & -47.519139 &  6 &-1.58 $\pm$ 0.10 &  0.12 $\pm$  0.01 & 13.70 $\pm$  0.02 &  \ldots          &  \ldots           &  \ldots           &  -1.85 $\pm$ 0.09 &  0.14 $\pm$  0.01 & 13.71 $\pm$  0.02  \\
 V146  &  0 & 201.720167 & -47.491194 &  6 & \ldots          &  \ldots           &  \ldots           & -2.33 $\pm$ 0.59 &  0.14 $\pm$  0.02 & 13.79 $\pm$  0.11 &  -1.74 $\pm$ 0.19 &  0.13 $\pm$  0.01 & 13.71 $\pm$  0.04  \\
 V147  &  1 & 201.816208 & -47.519528 &  6 &-2.13 $\pm$ 0.11 &  0.10 $\pm$  0.01 & 13.74 $\pm$  0.02 &  \ldots          &  \ldots           &  \ldots           &  -2.35 $\pm$ 0.17 &  0.12 $\pm$  0.01 & 13.72 $\pm$  0.03  \\
 V149  &  0 & 201.886875 & -47.228722 &  6 & \ldots          &  \ldots           &  \ldots           & -2.06 $\pm$ 0.13 &  0.14 $\pm$  0.01 & 13.72 $\pm$  0.03 &  -1.98 $\pm$ 0.09 &  0.14 $\pm$  0.01 & 13.72 $\pm$  0.02  \\
 V150  &  0 & 201.917625 & -47.600139 &  6 & \ldots          &  \ldots           &  \ldots           & -1.55 $\pm$ 0.49 &  0.07 $\pm$  0.01 & 13.66 $\pm$  0.09 &  -2.00 $\pm$ 0.18 &  0.07 $\pm$  0.01 & 13.73 $\pm$  0.04  \\
 V153  &  1 & 201.706875 & -47.440000 &  6 &-1.93 $\pm$ 0.16 &  0.15 $\pm$  0.01 & 13.72 $\pm$  0.03 &  \ldots          &  \ldots           &  \ldots           &  -1.99 $\pm$ 0.06 &  0.16 $\pm$  0.01 & 13.69 $\pm$  0.02  \\
 V154  &  1 & 201.763000 & -47.509222 &  6 &-2.00 $\pm$ 0.12 &  0.13 $\pm$  0.01 & 13.70 $\pm$  0.02 &  \ldots          &  \ldots           &  \ldots           &  -1.96 $\pm$ 0.14 &  0.13 $\pm$  0.01 & 13.71 $\pm$  0.03  \\
 V155  &  1 & 201.723500 & -47.411917 &  6 &-1.86 $\pm$ 0.09 &  0.12 $\pm$  0.01 & 13.73 $\pm$  0.02 &  \ldots          &  \ldots           &  \ldots           &  -2.06 $\pm$ 0.05 &  0.13 $\pm$  0.01 & 13.71 $\pm$  0.02  \\
 V156  &  1 & 201.699833 & -47.531500 &  5 &-1.48 $\pm$ 0.58 &  0.12 $\pm$  0.02 & 13.63 $\pm$  0.11 &  \ldots          &  \ldots           &  \ldots           &  -2.12 $\pm$ 0.10 &  0.14 $\pm$  0.01 & 13.72 $\pm$  0.06  \\
 V157  &  1 & 201.693583 & -47.454944 &  6 &-1.85 $\pm$ 0.26 &  0.14 $\pm$  0.01 & 13.73 $\pm$  0.04 &  \ldots          &  \ldots           &  \ldots           &  -2.03 $\pm$ 0.20 &  0.16 $\pm$  0.01 & 13.71 $\pm$  0.04  \\
 V158  &  1 & 201.688792 & -47.511250 &  6 &-1.93 $\pm$ 0.21 &  0.11 $\pm$  0.01 & 13.75 $\pm$  0.03 &  \ldots          &  \ldots           &  \ldots           &  -1.88 $\pm$ 0.13 &  0.12 $\pm$  0.01 & 13.72 $\pm$  0.03  \\
 V160  &  1 & 201.400292 & -47.208972 &  6 &-1.80 $\pm$ 0.16 &  0.10 $\pm$  0.01 & 13.75 $\pm$  0.03 &  \ldots          &  \ldots           &  \ldots           &  -1.91 $\pm$ 0.11 &  0.11 $\pm$  0.01 & 13.72 $\pm$  0.03  \\
 V163  &  1 & 201.456125 & -47.339389 &  6 &-1.82 $\pm$ 0.23 &  0.11 $\pm$  0.01 & 13.71 $\pm$  0.04 &  \ldots          &  \ldots           &  \ldots           &  -1.88 $\pm$ 0.18 &  0.11 $\pm$  0.01 & 13.74 $\pm$  0.03  \\
 V165  &  0 & 201.664083 & -47.448889 &  3 & \ldots          &  \ldots           &  \ldots           & -2.31 $\pm$ 0.83 &  0.12 $\pm$  0.09 & 13.87 $\pm$  0.17 &  -1.43 $\pm$ 0.10 &  0.18 $\pm$  0.06 & 13.69 $\pm$  0.04  \\
 V169  &  1 & 201.835208 & -47.399917 &  6 &-1.92 $\pm$ 0.18 &  0.12 $\pm$  0.01 & 13.74 $\pm$  0.03 &  \ldots          &  \ldots           &  \ldots           &  -1.72 $\pm$ 0.11 &  0.11 $\pm$  0.01 & 13.73 $\pm$  0.03  \\
 V184  &  1 & 201.868708 & -47.526694 &  6 &-1.87 $\pm$ 0.24 &  0.14 $\pm$  0.01 & 13.68 $\pm$  0.04 &  \ldots          &  \ldots           &  \ldots           &  -1.93 $\pm$ 0.14 &  0.13 $\pm$  0.01 & 13.73 $\pm$  0.03  \\
 V185  &  1 & 201.516917 & -47.363056 &  6 &-2.04 $\pm$ 0.29 &  0.08 $\pm$  0.01 & 13.78 $\pm$  0.04 &  \ldots          &  \ldots           &  \ldots           &  -1.80 $\pm$ 0.18 &  0.08 $\pm$  0.01 & 13.75 $\pm$  0.03  \\
 V261  &  1 & 201.814167 & -47.358333 &  6 &-1.98 $\pm$ 0.85 &  0.20 $\pm$  0.02 & 13.55 $\pm$  0.11 &  \ldots          &  \ldots           &  \ldots           &  -2.91 $\pm$ 0.03 &  0.22 $\pm$  0.01 & 13.66 $\pm$  0.02  \\
 V263  &  0 & 201.554625 & -47.436194 &  6 & \ldots          &  \ldots           &  \ldots           & -1.45 $\pm$ 0.50 &  0.08 $\pm$  0.01 & 13.67 $\pm$  0.08 &  -1.95 $\pm$ 0.08 &  0.09 $\pm$  0.01 & 13.75 $\pm$  0.02  \\
 V264  &  1 & 201.665167 & -47.507944 &  6 &-1.81 $\pm$ 0.43 &  0.15 $\pm$  0.01 & 13.76 $\pm$  0.06 &  \ldots          &  \ldots           &  \ldots           &  -1.42 $\pm$ 0.04 &  0.14 $\pm$  0.01 & 13.71 $\pm$  0.02  \\
 V265  &  1 & 201.625833 & -47.479361 &  6 &-1.84 $\pm$ 0.13 &  0.11 $\pm$  0.01 & 13.71 $\pm$  0.02 &  \ldots          &  \ldots           &  \ldots           &  -2.24 $\pm$ 0.05 &  0.13 $\pm$  0.01 & 13.71 $\pm$  0.02  \\
 V266  &  1 & 201.665042 & -47.467250 &  6 &-1.82 $\pm$ 0.13 &  0.11 $\pm$  0.01 & 13.71 $\pm$  0.02 &  \ldots          &  \ldots           &  \ldots           &  -1.96 $\pm$ 0.13 &  0.12 $\pm$  0.01 & 13.72 $\pm$  0.03  \\
 V267  &  1 & 201.667458 & -47.443361 &  6 &-2.41 $\pm$ 0.20 &  0.12 $\pm$  0.01 & 13.73 $\pm$  0.04 &  \ldots          &  \ldots           &  \ldots           &  -2.11 $\pm$ 0.10 &  0.11 $\pm$  0.01 & 13.72 $\pm$  0.03  \\
 V268  &  0 & 201.646292 & -47.436472 &  5 & \ldots          &  \ldots           &  \ldots           & -1.91 $\pm$ 0.17 &  0.15 $\pm$  0.01 & 13.68 $\pm$  0.10 &  -2.07 $\pm$ 0.02 &  0.15 $\pm$  0.01 & 13.71 $\pm$  0.09  \\
 V270  &  1 & 201.735542 & -47.501722 &  6 &-1.96 $\pm$ 0.18 &  0.12 $\pm$  0.01 & 13.69 $\pm$  0.03 &  \ldots          &  \ldots           &  \ldots           &  -1.99 $\pm$ 0.14 &  0.11 $\pm$  0.01 & 13.72 $\pm$  0.03  \\
 V271  &  1 & 201.696250 & -47.501194 &  6 &-1.83 $\pm$ 0.14 &  0.12 $\pm$  0.02 & 13.74 $\pm$  0.03 &  \ldots          &  \ldots           &  \ldots           &  -2.10 $\pm$ 0.14 &  0.15 $\pm$  0.02 & 13.71 $\pm$  0.03  \\
 V272  &  1 & 201.678792 & -47.432472 &  6 &-2.10 $\pm$ 0.19 &  0.15 $\pm$  0.01 & 13.68 $\pm$  0.03 &  \ldots          &  \ldots           &  \ldots           &  -2.10 $\pm$ 0.14 &  0.15 $\pm$  0.01 & 13.71 $\pm$  0.03  \\
 V273  &  1 & 201.726375 & -47.452528 &  6 &-1.84 $\pm$ 0.05 &  0.13 $\pm$  0.01 & 13.71 $\pm$  0.02 &  \ldots          &  \ldots           &  \ldots           &  -2.00 $\pm$ 0.08 &  0.14 $\pm$  0.01 & 13.71 $\pm$  0.02  \\
 V274  &  1 & 201.682167 & -47.380111 &  6 &-1.71 $\pm$ 0.32 &  0.12 $\pm$  0.01 & 13.68 $\pm$  0.05 &  \ldots          &  \ldots           &  \ldots           &  -1.89 $\pm$ 0.16 &  0.12 $\pm$  0.01 & 13.73 $\pm$  0.03  \\
 V275  &  1 & 201.707208 & -47.460417 &  6 &-2.02 $\pm$ 0.10 &  0.13 $\pm$  0.01 & 13.72 $\pm$  0.02 &  \ldots          &  \ldots           &  \ldots           &  -2.08 $\pm$ 0.10 &  0.14 $\pm$  0.01 & 13.70 $\pm$  0.03  \\
 V276  &  1 & 201.818708 & -47.555056 &  6 &-2.08 $\pm$ 0.19 &  0.14 $\pm$  0.01 & 13.69 $\pm$  0.03 &  \ldots          &  \ldots           &  \ldots           &  -2.07 $\pm$ 0.14 &  0.13 $\pm$  0.01 & 13.72 $\pm$  0.03  \\
 V277  &  1 & 201.749792 & -47.458250 &  6 &-2.16 $\pm$ 0.19 &  0.11 $\pm$  0.01 & 13.74 $\pm$  0.03 &  \ldots          &  \ldots           &  \ldots           &  -2.08 $\pm$ 0.16 &  0.11 $\pm$  0.01 & 13.73 $\pm$  0.03  \\
 V280  &  1 & 201.788875 & -47.385083 &  6 &-1.71 $\pm$ 0.19 &  0.13 $\pm$  0.01 & 13.73 $\pm$  0.03 &  \ldots          &  \ldots           &  \ldots           &  -1.41 $\pm$ 0.12 &  0.11 $\pm$  0.01 & 13.73 $\pm$  0.02  \\
 V285  &  1 & 201.417083 & -47.580167 &  6 &-1.77 $\pm$ 0.15 &  0.11 $\pm$  0.01 & 13.73 $\pm$  0.03 &  \ldots          &  \ldots           &  \ldots           &  -1.76 $\pm$ 0.17 &  0.11 $\pm$  0.01 & 13.74 $\pm$  0.03  \\
 V288  &  1 & 202.043333 & -47.396556 &  6 &-2.33 $\pm$ 0.37 &  0.14 $\pm$  0.01 & 13.75 $\pm$  0.06 &  \ldots          &  \ldots           &  \ldots           &  -1.84 $\pm$ 0.17 &  0.12 $\pm$  0.01 & 13.72 $\pm$  0.03  \\
 V289  &  1 & 202.014792 & -47.357778 &  6 &-2.06 $\pm$ 0.14 &  0.15 $\pm$  0.01 & 13.70 $\pm$  0.02 &  \ldots          &  \ldots           &  \ldots           &  -1.93 $\pm$ 0.15 &  0.14 $\pm$  0.01 & 13.71 $\pm$  0.03  \\
 V291  &  1 & 201.660500 & -47.557889 &  6 &-1.55 $\pm$ 0.61 &  0.08 $\pm$  0.02 & 13.66 $\pm$  0.08 &  \ldots          &  \ldots           &  \ldots           &  -2.11 $\pm$ 0.17 &  0.10 $\pm$  0.01 & 13.75 $\pm$  0.03  \\
 NV339 &  1 & 201.623583 & -47.497889 &  6 &-2.31 $\pm$ 0.47 &  0.14 $\pm$  0.02 & 13.63 $\pm$  0.07 &  \ldots          &  \ldots           &  \ldots           &  -2.49 $\pm$ 0.18 &  0.14 $\pm$  0.01 & 13.70 $\pm$  0.04  \\
 NV340 &  1 & 201.662125 & -47.459167 &  6 &-1.45 $\pm$ 0.82 &  0.12 $\pm$  0.02 & 13.59 $\pm$  0.10 &  \ldots          &  \ldots           &  \ldots           &  -2.08 $\pm$ 0.11 &  0.13 $\pm$  0.01 & 13.72 $\pm$  0.02  \\
 NV341 &  1 & 201.727625 & -47.480139 &  6 &-1.91 $\pm$ 0.49 &  0.13 $\pm$  0.02 & 13.63 $\pm$  0.07 &  \ldots          &  \ldots           &  \ldots           &  -2.21 $\pm$ 0.11 &  0.13 $\pm$  0.01 & 13.71 $\pm$  0.03  \\
 NV343 &  1 & 201.699125 & -47.493778 &  6 &-2.04 $\pm$ 0.29 &  0.14 $\pm$  0.01 & 13.66 $\pm$  0.04 &  \ldots          &  \ldots           &  \ldots           &  -2.12 $\pm$ 0.19 &  0.13 $\pm$  0.01 & 13.70 $\pm$  0.04  \\
 NV344 &  1 & 201.658458 & -47.412500 &  6 &-2.23 $\pm$ 0.28 &  0.12 $\pm$  0.01 & 13.75 $\pm$  0.05 &  \ldots          &  \ldots           &  \ldots           &  -1.90 $\pm$ 0.17 &  0.11 $\pm$  0.01 & 13.73 $\pm$  0.04  \\
 NV346 &  1 & 201.695500 & -47.470667 &  6 &-1.60 $\pm$ 0.67 &  0.11 $\pm$  0.02 & 13.61 $\pm$  0.09 &  \ldots          &  \ldots           &  \ldots           &  -2.18 $\pm$ 0.12 &  0.12 $\pm$  0.01 & 13.71 $\pm$  0.03  \\
 NV347 &  1 & 201.574500 & -47.483194 &  6 &-1.40 $\pm$ 1.40 &  0.14 $\pm$  0.03 & 13.48 $\pm$  0.15 &  \ldots          &  \ldots           &  \ldots           &  -2.63 $\pm$ 0.19 &  0.16 $\pm$  0.01 & 13.67 $\pm$  0.04  \\
 NV349 &  1 & 201.715792 & -47.462306 &  6 &-2.10 $\pm$ 0.57 &  0.12 $\pm$  0.02 & 13.64 $\pm$  0.09 &  \ldots          &  \ldots           &  \ldots           &  -2.64 $\pm$ 0.17 &  0.13 $\pm$  0.02 & 13.71 $\pm$  0.05  \\
 NV350 &  1 & 201.734875 & -47.514083 &  6 &-1.57 $\pm$ 0.42 &  0.11 $\pm$  0.01 & 13.66 $\pm$  0.06 &  \ldots          &  \ldots           &  \ldots           &  -2.13 $\pm$ 0.12 &  0.13 $\pm$  0.01 & 13.71 $\pm$  0.03  \\
 NV352 &  1 & 201.726583 & -47.486722 &  6 &-2.16 $\pm$ 0.62 &  0.13 $\pm$  0.02 & 13.64 $\pm$  0.09 &  \ldots          &  \ldots           &  \ldots           &  -2.81 $\pm$ 0.19 &  0.15 $\pm$  0.01 & 13.71 $\pm$  0.04  \\
 NV353 &  1 & 201.682417 & -47.465806 &  6 &-2.05 $\pm$ 0.61 &  0.14 $\pm$  0.02 & 13.62 $\pm$  0.09 &  \ldots          &  \ldots           &  \ldots           &  -2.70 $\pm$ 0.23 &  0.17 $\pm$  0.01 & 13.68 $\pm$  0.05  \\
 NV354 &  1 & 201.660750 & -47.419528 &  6 &-1.78 $\pm$ 0.10 &  0.11 $\pm$  0.01 & 13.72 $\pm$  0.02 &  \ldots          &  \ldots           &  \ldots           &  -2.13 $\pm$ 0.09 &  0.14 $\pm$  0.01 & 13.71 $\pm$  0.02  \\
 NV357 &  1 & 201.573958 & -47.506694 &  6 &-2.04 $\pm$ 0.24 &  0.14 $\pm$  0.01 & 13.67 $\pm$  0.04 &  \ldots          &  \ldots           &  \ldots           &  -2.07 $\pm$ 0.11 &  0.14 $\pm$  0.01 & 13.72 $\pm$  0.03  \\
 NV366 &  0 & 201.673125 & -47.528444 &  6 & \ldots          &  \ldots           &  \ldots           & -2.04 $\pm$ 0.94 &  0.07 $\pm$  0.02 & 13.59 $\pm$  0.15 &  -2.99 $\pm$ 0.17 &  0.09 $\pm$  0.01 & 13.74 $\pm$  0.03  \\
 NV399 &  1 & 201.623000 & -47.500889 &  6 &-1.49 $\pm$ 0.42 &  0.12 $\pm$  0.01 & 13.67 $\pm$  0.06 &  \ldots          &  \ldots           &  \ldots           &  -1.81 $\pm$ 0.13 &  0.12 $\pm$  0.01 & 13.73 $\pm$  0.03  \\
\enddata
\tablenotetext{a}{~Varible identification according to \citet{braga16}.}
\tablenotetext{b}{~Pulsation mode: 0 -- RRab; 1 -- RRc.} 
\tablenotetext{c}{~Right ascension and declination according to \citet{braga16}.}
\tablenotetext{d}{~Number of mean magnitudes adopted for the REDIME solution.}
\tablenotetext{e}{~Parameter estimates based on RRc, RRab and Global REDIME solution.}
\tablenotetext{f}{~Mean iron abundance and standard deviation based on NIR ($J$,$H$,$K$) PL relations.}
\tablenotetext{g}{~Reddening and its uncertainty.}
\tablenotetext{h}{~True distance modulus and its uncertainty.}
\label{tab:fe_mi_ebv}
\end{deluxetable*}

\clearpage 
\setlength{\tabcolsep}{0.15cm}
\begin{deluxetable*}{lcccccc}
 \tablewidth{0pt}
 \tabletypesize{\scriptsize}
 \tablecaption{Mean (metal abundance) and median (true distance modulus, 
reddening) estimates based on REDIME (first and second iteration) for 
RRc, RRab and Global solution. The errors indicate the error on the 
mean/median and their standard deviation.}
 \tablehead{ & & \colhead{First Iteration} & & & \colhead{Second Iteration} & \\
 \colhead{} & \colhead{[Fe/H]} & \colhead{$\mu$} & \colhead{E($B-V$)}  & \colhead{[Fe/H]} & \colhead{$\mu$} & \colhead{E($B-V$)} \\
 & dex & mag & mag & dex & mag & mag}
  \startdata
RRc  &    -1.98 $\pm$ 0.06 $\pm$ 0.53    &  13.722 $\pm$ 0.006 $\pm$ 0.061  &   0.124 $\pm$ 0.013 $\pm$ 0.028 & -1.97 $\pm$ 0.06 $\pm$ 0.54 & 13.721 $\pm$ 0.003 $\pm$ 0.029 & 0.128 $\pm$ 0.002 $\pm$ 0.025 \\
RRab &    -1.93 $\pm$ 0.06 $\pm$ 0.53    &  13.711 $\pm$ 0.004 $\pm$ 0.042  &   0.140 $\pm$ 0.016 $\pm$ 0.030 & -1.91 $\pm$ 0.06 $\pm$ 0.56 & 13.718 $\pm$ 0.003 $\pm$ 0.030 & 0.136 $\pm$ 0.003 $\pm$ 0.030 \\
Global &  -1.96 $\pm$ 0.04 $\pm$ 0.52    &  13.717 $\pm$ 0.005 $\pm$ 0.071  &   0.131 $\pm$ 0.010 $\pm$ 0.027 & -1.98 $\pm$ 0.04 $\pm$ 0.54 & 13.720 $\pm$ 0.002 $\pm$ 0.030 & 0.132 $\pm$ 0.002 $\pm$ 0.028 \\
%
\enddata
\label{tab2}
\end{deluxetable*}

%
\setlength{\tabcolsep}{0.15cm}
\begin{deluxetable}{l l l l}
 \tablewidth{0pt}
 \tabletypesize{\scriptsize}
 \tablecaption{True distance moduli and reddening for $\omega$~Cen available 
in the literature.}
 \tablehead{\colhead{$\mu$} & \colhead{E($B-V$)}\tablenotemark{a}  & \colhead{Reference} & Notes\tablenotemark{b} \\ 
 ~~~~~~~~~mag   & ~~~~mag    &  & }  
\startdata
13.36  $\pm$  0.10  &  0.11\tablenotemark{c}   &  \cite{cannon1974}  & (1) \\
13.61               &  0.11  \citep{buonanno89b}  & \cite{longmore1990}  & (2) \\
13.53  $\pm$  0.20  &  0.11\tablenotemark{d}   &  \cite{nemec1994}  & (3) \\ 
14.02  $\pm$  0.10  &  0.12\tablenotemark{d}   &  \cite{mcnamara00} & (4) \\
13.74  $\pm$  0.11  &  0.11\tablenotemark{c}   &  \cite{caputo2002b}  & (5) \\
13.75  $\pm$  0.04  &  0.11\tablenotemark{c}   & \cite{kaluzny2002} & (6) \\
13.72  $\pm$  0.11  &  0.11$\pm$0.01 \citep{calamida05} & \cite{delprincipe06}  & (7) \\
13.62  $\pm$  0.11  &  0.11$\pm$0.01 \citep{calamida05} & \cite{delprincipe06}  & (7) \\
13.77  $\pm$  0.07  &  0.11$\pm$0.01 \citep{calamida05} & \cite{delprincipe06}  & (8) \\
13.72               &  0.11$\pm$0.01 \citep{ferraro1999} & \cite{sollima06a}  & (9) \\
13.49  $\pm$  0.14  &  0.13 \citep{schlegel98}  & \citep{kaluzny07a}  & (10) \\
13.51  $\pm$  0.12  &  0.13 \citep{schlegel98}  & \citep{kaluzny07a}  & (10) \\
13.68  $\pm$  0.27  &  0.12 \citep{harris96}  &  \cite{weldrake2007}  & (11) \\ 
13.65  $\pm$  0.09  &  0.11$\pm$0.02 \citep{calamida05} & \cite{bono2008b}  & (12) \\
13.75  $\pm$  0.11  &  0.11$\pm$0.02 \citep{calamida05} & \cite{bono2008b}  & (13) \\
13.62  $\pm$  0.05  &  {\bf 0.16}    &  \cite{mcnamara11}  & (14) \\
13.71  $\pm$  0.08  &  0.11  \citep{thompson2001,lub2002}  &  \cite{braga16}  & (15) \\
13.65  $\pm$  0.08  &  0.12 \citep{harris96}  & \cite{bhardwaj17b}  & (16) \\ 
13.77  $\pm$  0.08  &  0.12 \citep{harris96}  & \cite{bhardwaj17b}  & (16) \\ 
13.70  $\pm$  0.11  &  0.12 \citep{harris96}  & \cite{bhardwaj17b}  & (16) \\ 
13.708 $\pm$  0.035 &  0.12  \citep{harris96}  & \cite{navarrete17}  & (17) \\ 
13.674 $\pm$  0.038 &  0.11 \citep{thompson2001,lub2002} & \cite{braga2018}  & (18) \\
13.698 $\pm$  0.048 &  0.11 \citep{thompson2001,lub2002} & \cite{braga2018}  & (18) \\
13.720 $\pm$  0.002 $\pm$ 0.030 &  {\bf 0.132 $\pm$ 0.002 $\pm$ 0.028}   & This work  & (19) \\
\enddata
\tablenotetext{a}{The reddening estimates, E(\bmv),  derived by the authors of the 
investigation are marked in bold. For the investigations in which the reddening was 
assumed from the literature we include the reference.}
\tablenotetext{b}{
(1) Distance based on the visual band, $M_V$, of the Horizontal Branch.
(2) Distance based on the $K$-band PL relation.  The relation was calibrated by using  
$M_{K,o,-0.3} = 0.06$ and [Fe/H] = --0.24. 
(3) Distance based on the $B_{0,-0.3}$, $V_{0,-0.3}$ and $K_{0,-0.3}$ magnitudes 
of RRLs, where the subscript $0,-0.3$ means the reddening-corrected 
magnitude at $\log{P} = -0.3$.
(4) Distance based on the $V$-band PL relation of high-amplitude $\delta$ Sct stars.
(5) Distnce based on the position of the First Overtone Blue Edge of the instability strip
in the $\log P$-$M_V$ diagram.
(6) Distance based on the surface brightness method, applied to the detached 
eclipsing binary V212 to derive the absolute distance to $\omega$ Cen.
(7) Distance based on the $M_V$-[Fe/H] relation, calibrated with \citet{bono03c} and 
with \citet{Catelan06} for the two values. 
(8) Distance based on the semi-empirical calibration of the $K_s$-band PL relation 
by \citet{bono03c}.
(9) Distance based on a new calibration of the $K_s$-band PL relation. The zero-point 
was based on the trigonometric parallax of the prototype RR Lyr \citep{benedict11}.
(10) Distance based on the orbital parameters of the detached eclipsing binary 
V209. The two distance moduli are for the primary (closest) and for the secondary 
(farthest) star of the binary system. 
(11) Distance based on the $M_V$-[Fe/H] relation, calibrated with \citet{rich05}.
(12) Distance based on the calibration of the TRGB provided by \citet{lee93}.
(13) Distance based on the empirical $K$-band PL relation provided by \citep{sollima2008}.
(14) Distance based on the $V$-band PL relation of $\delta$ Sct stars.
(15) Distance based on semi-empirical and theoretical calibration of the reddening 
independent PW($V$,\bmi) relations. 
(16) Distance based on the the $J$-, $H$- and $K$-band PL relations of 
Type II Cepheids (T2Cs), based on a new calibration of the 
Large Magellanic Cloud Type II Cepheids.
(17) Distance based on the $J$- and $K$-band PL relations of both RRLs and 
Type II Cepheids, calibrated with the relations of \citep{alonsogarcia2015}.
(18) Distance based on the $J$-, $H$- and $K_s$-band PLZ relations of RRLs, 
calibrated with the predicted relations and adopting [Fe/H] from
\citet{sollima06a} and \citet{braga16}. E(\bmv) from \citep{thompson2001,lub2002}.
(19) Distance and reddening based on the application of REDIME to
the $BVIJHK_s$ mean magnitudes of RRLs.}
\tablenotetext{c}{The authors provide apparent distance modulus ($m-M$)$_V$ 
and not the true distance modulus $\mu$. Therefore, we adopt E$(B-V)$=0.11 
\citep{thompson2001,lub2002} and provide $\mu$ in column 1.}
\tablenotetext{d}{The authors do not quote the paper from which the reddening 
value was adopted.}
\label{tab3}
\end{deluxetable}


\end{document}